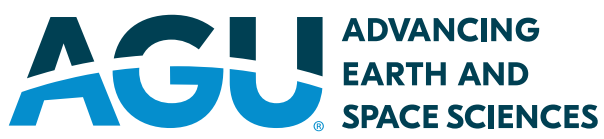

**Key Points:**

- During the late stage of the substorm recovery phase, global-scale transpolar arc and local-scale auroral spiral can appear simultaneously
- The magnetotail field-aligned current (FAC) intensity of the auroral spiral was about 3 orders of magnitude weaker than that of the TPA
- Weak spiral FAC formation appears to require solar wind–magnetosphere–ionosphere coupling with minimal or no substorm activity

**Supporting Information:**

Supporting Information may be found in the online version of this article.

**Correspondence to:**

M. Nowada,
moto.nowada@sdu.edu.cn;
nowada.sdu@gmail.com

**Author Contributions:**

**Conceptualization:** Motoharu Nowada
**Data curation:** Motoharu Nowada, Yukinaga Miyashita, Aoi Nakamizo, Noora Partamies
**Formal analysis:** Motoharu Nowada, Yukinaga Miyashita, Aoi Nakamizo, Noora Partamies
**Funding acquisition:** Motoharu Nowada, Quan-Qi Shi
**Investigation:** Motoharu Nowada, Yukinaga Miyashita, Aoi Nakamizo, Noora Partamies
**Methodology:** Motoharu Nowada

# Contemporaneous Appearances of Auroral Spiral and Transpolar Arc: Polar UVI Observations and Global MHD Simulations

**Motoharu Nowada**[1], **Yukinaga Miyashita**[2,3], **Aoi Nakamizo**[4], **Noora Partamies**[5], and **Quan-Qi Shi**[1]

[1]Shandong Key Laboratory of Optical Astronomy and Solar-Terrestrial Environment, School of Space Science and Physics, Institute of Space Sciences, Shandong University, Weihai, China, [2]Center for Heliophysics Research, Korea Astronomy and Space Science Institute, Daejeon, South Korea, [3]Department of Astronomy and Space Science, Korea University of Science and Technology, Daejeon, South Korea, [4]National Institute of Communication and Technology, Koganei, Japan, [5]Department of Arctic Geophysics, The University Centre in Svalbard, Longyearbyen, Norway

**Abstract** A local vortex-structured aurora and a large-scale transpolar arc (TPA) were contemporaneously observed by the Polar ultraviolet imager (UVI) during the late recovery phase of a substorm, and the interplanetary magnetic field (IMF) $B_Y$ and $B_Z$ were negative and negative-to-positive. The TPA grew along the dawnside auroral oval from the nightside to the dayside, and an auroral spiral and several spots were located azimuthally near the poleward edge of the nightside auroral oval. Both auroras had tailward elongated source regions with scales of ∼30 $R_E$ (spiral) and more than ∼45 $R_E$ (TPA). To examine their magnetospheric/ionospheric field-aligned current (FAC) profiles, we performed global magnetohydrodynamic (MHD) simulations, using two different types of code: Block-Adaptive-Tree Solar-wind Roe Upwind Scheme (BATS-R-US) and improved REProduce Plasma Universe (REPPU). Both MHD simulations reproduced the tailward elongated TPA-associated FAC structures. The spiral-associated FAC intensity was, however, approximately three orders of magnitude weaker than the TPA-associated FAC intensity. Only improved REPPU simulations replicated faint but continuous poleward extending streak-like structures without evident FACs, instead of the auroral spiral. Geomagnetic field measurements showed that the spiral had upward (from the ionosphere to the magnetosphere) FACs, and its appearance might be accompanied by ultra-low-frequency Pc5 waves. Our results suggest that (a) a local-scale spiral might be formed with much weaker magnetotail FACs than global-scale TPA-associated FACs, although the spiral source region is elongated tailward, and (b) a solar wind-magnetosphere-ionosphere coupling system with minimal or no significant substorm effects is required to form the spiral with the weak magnetotail FACs.

**Plain Language Summary** Based on in situ satellite and ground observations, combined with global magnetohydrodynamic MHD computer simulations, recent studies have examined the auroral spiral, local-scale vortex-structured auroral phenomenon, and global-scale transpolar arc (TPA) bridging the polar cap between the nightside and dayside sectors of the auroral oval. These investigations advance our understanding of how magnetotail processes contribute to such formations within the solar wind–magnetosphere–ionosphere coupling system, particularly under conditions minimally influenced by substorms. This study reports the contemporaneous appearance of an auroral spiral and a TPA in the polar cap. Using global MHD simulations with two different types of codes, we successfully reproduced this concomitant occurrence and investigated the corresponding field-aligned current (FAC) structures in the nightside magnetospheric equatorial plane. Both simulations revealed that (a) the FACs associated with TPA exhibited tailward-elongated structures, and (b) the spiral-associated FAC intensity was approximately three orders of magnitude weaker than the TPA-associated FAC intensity. Furthermore, ground-based geomagnetic field observations suggest that ultra-low-frequency MHD waves might play a role in the auroral spiral formation. To explain how local-scale auroral spiral form and its associated FACs effectively reach the polar cap, a new solar wind-magnetosphere-ionosphere coupling system with minimum substorm effects is required.

## 1. Introduction

Auroral phenomena that locally exhibit a vortex structure and mainly appear as azimuthally-aligned spot emissions in the poleward region of the nightside auroral oval are referred to as auroral spirals (e.g., Elphinstone et al., 1995; Nowada et al., 2023; Partamies et al., 2001a). A large number of auroral spirals were statistically observed over the nightside ionosphere from 18 to 5 h magnetic local time (MLT), with their occurrences predominantly concentrated around 65° magnetic latitude (MLat; Partamies et al., 2001a) and between 70° and 80° MLat (Davis & Hallinan, 1976). Statistical diameter distributions of auroral spirals are generally 25–75 km (Partamies et al., 2001a, 2001b), or 20–1,300 km (Davis & Hallinan, 1976). Recently, it has been reported that the auroral spiral can appear during various substorm phases: expansion phase (Haerendel & Partamies, 2024; Keiling, Angelopoulos, Runov, et al., 2009; Keiling, Angelopoulos, Weygand, et al., 2009) and late substorm recovery phase (Nowada et al., 2023). Depending on the substorm phase of spiral occurrence, the fundamental spiral features, such as scale, duration, and formation mechanism, are expected to differ (e.g., Nowada et al., 2023). These differences can be attributed to the distinct magnetotail configuration and the associated plasma dynamics that characterize the substorm expansion and recovery phases. Consequently, we consider that the spiral formation process (mechanisms) and its source regions are believed to be fundamentally distinct between two substorm phases. Keiling, Angelopoulos, Runov, et al. (2009) and Keiling, Angelopoulos, Weygand, et al. (2009) examined auroral spirals that occurred from the substorm onset to the expansion phase, and estimated its scale at 200–300 km. In contrast, the auroral spiral observed during the late stage of the substorm recovery phase had smaller core scale from 150 to 250 km (Nowada et al., 2023).

In the early days of auroral spiral studies, Davis and Hallinan (1976) and Hallinan and Davis (1970) showed that in addition to the spiral, curl- and fold-type auroras were accompanied by vortex structures. Spirals are the auroral phenomena with the largest vortex whose scale is larger than 50 km. Curls are small-scale auroras with vortices of less than 10 km, which are embedded in auroral arcs. Folds are distorted arcs with an intermediate scale of about 20 km. These vortical auroras, however, can appear contemporaneously in the same auroral form and could not be easily distinguished. Recently, based Polar ultraviolet imager (UVI) and all-sky camera imager data, Nowada et al. (2023) followed the auroral spiral formation process after the substorm-associated auroral bulge in the nightside ionosphere subsided. After the substorm entered the recovery phase, several poleward-elongated auroral streak-like structures appeared in the nightside auroral oval. Finally, the auroral spiral was formed during the late stage of the substorm recovery phase. Nowada et al. (2023) also estimated the source region of the auroral spiral on the nightside magnetotail equatorial plane using an empirical magnetic field model. Their results demonstrated that the auroral spiral source region on the nightside equatorial plane extended ~30 $R_E$ tailward and ~4 $R_E$ in the dawn-dusk direction. They, however, could not physically reveal how the spiral was formed, based on the relations with the solar wind conditions and the corresponding magnetotail processes.

At present, the formation mechanism of the auroral spirals remains unclear. Keiling, Angelopoulos, Weygand, et al. (2009) and Voronkov et al. (2000) proposed that the spiral might be formed by shear flow ballooning instability which is treated within a magnetohydrodynamic (MHD) regime and is driven by a pressure gradient in the magnetotail. Furthermore, the auroral spiral formation cannot be yet addressed from the viewpoint of solar wind-magnetosphere-ionosphere coupling system. Lysak and Song (1996), however, proposed a spiral formation model based on magnetosphere-ionosphere coupling within a magnetohydrodynamics (MHD) scheme. The current sheet instability (CSI) in the nightside ionosphere can play a crucial role in formation of the auroral spiral. Current sheet instability is quite similar to the Kelvin-Helmholtz instability (KHI) caused by a velocity shear, but a magnetic shear generated by an upward (from the ionosphere to the magnetosphere) field-aligned current (FAC) plays a significant role in its growth. Lysak and Song (1996) concluded that an auroral spiral can be formed by CSI because of the conductance difference between the ionosphere and the magnetosphere. Hallinan (1976) and Partamies et al. (2001b) also pointed out that enhancement of an upward FAC filament structure drives the magnetic shear and the resulting CSI to form the auroral spiral. Although the two groups of Lysak and Song (1996), and Hallinan (1976) and Partamies et al. (2001b) independently proposed the models to explain the auroral spiral formation, they all emphasized that the CSI due to an upward FAC is the major mechanism of the auroral spiral formation. At present, however, we do not have the observational evidence for their models and the CSI reproductions based on global-scale computer simulations using MHD code, suggesting that the auroral spiral features, including the formation process, cannot yet be explained completely within the MHD regime.

Similar vortex-shaped auroral structures reported in Saturn's magnetosphere (Radioti et al., 2015) imply that the underlying processes may be universal across planetary environments. While this comparative perspective is valuable, the present work focuses on Earth, where the simultaneous occurrence of a spiral and a TPA provides new constraints on magnetotail–ionosphere coupling.

Transpolar arc (TPA) is a "crossbar"-type auroral phenomenon bridging the polar cap from the nightside to the dayside auroral oval and is also a part of $\theta$-like-shaped aurora (theta aurora). The formation process and fundamental physical features of TPA have been investigated and discussed (see reviews by Fear & Milan, 2012a, 2012b) since a theta aurora was detected by the Dynamics Explorer (DE) 1 spacecraft (Frank et al., 1982). As a representative TPA formation model, Milan et al. (2005) built a formation model based on nightside magnetic reconnection, which successfully explained the formations in several TPA cases (Fear & Milan, 2012a, 2012b; Kullen et al., 2015; Nowada et al., 2018, and references therein). The development of TPA from the nightside to the dayside poleward edge of the auroral oval can be explained by continuous formation of newly closed field lines generated by nightside reconnection retreating its site toward the farther tail. Nowada et al. (2020) proposed a possible formation model of the dawnside (duskside) TPA and concluded that the source of the TPA is FACs induced by the plasma velocity shear between a fast plasma flow caused by nightside magnetic reconnection and the slower background magnetotail plasma flow. The TPA connects the nightside and dayside sectors of the auroral oval, but its formation requires magnetic field line twisting and magnetotail deformation due to the IMF-BY component, which had already been revealed by many researchers (e.g., Cowley, 1981, 1994; Gosling et al., 1990; Tsyganenko et al., 2015; Tsyganenko and Fairfield, 2004; Tsyganenko and Sitnov, 2005).

Many researches successfully reproduced TPAs, based on global MHD simulations (e.g., Kullen and Janhunen, 2004; Tanaka et al., 2004; Watanabe et al., 2014; and references therein). This suggests that an auroral phenomenon of TPA itself can be treated and discussed within an MHD framework. The TPA is actually identified as global-scale aurora because it connects dayside and nightside polar cap regions, and has significant conjugacy between the Northern and Southern Hemispheres (e.g., Carter et al., 2017; Nowada et al., 2020).

In this study, we examined a unique case observed by the Polar UVI on 10 January 1997 in which the auroral morphologies temporally and drastically changed before and after a substorm. Before the substorm growth phase, the TPAs aligned with the dawnside auroral oval were observed. Even though the substorm headed for complete recovery, a TPA appeared and remained, and even an auroral spiral appeared simultaneously. In particular, focusing on the interval of contemporaneous appearances of the local-scale auroral spiral and the global-scale TPA, we examined the magnetotail and ionospheric FAC profiles and structures on the nightside magnetic equatorial plane and in the polar cap. Furthermore, based on two different global MHD simulations, as well as the Polar UVI observations, we argued what physical condition the auroral spiral and the TPA can be concomitant.

The instrumentation and the two global MHD simulation codes (Space Weather Modeling Framework (SWMF) with BATS-R-US and REPPU codes) are described in Sections 2 and 3. In Section 4, we show the solar wind conditions, the results of the Polar UVI observations and global MHD simulations based on two different types of code, the comparison between the observations and the simulations, and geomagnetic field measurements. The discussions and conclusions of this study are described in Section 5.

## 2. Instrumentation of the Ultraviolet Auroral Imager

The ultraviolet imager (UVI) onboard Polar, which was launched on 24 February 1996, provides global auroral imaging data in ultraviolet range (Torr et al., 1995). We used the UVI images in altitude adjusted corrected geomagnetic (AACGM; Baker & Wing, 1989) and geographic coordinates. The UVI image data are degraded in the direction perpendicular to the track of Polar by the satellite's wobble motion (e.g., Parks et al., 1997). This wobble is, however, predictable (Parks et al., 1997), so we used the UVI image data from which the wobble effects were mostly removed.

## 3. Global MHD Simulation Codes

### 3.1. Space Weather Modeling Framework (SWMF) With Block-Adaptive-Tree Solar-Wind Roe Upwind Scheme (BATS-R-US) Model

The Space Weather Modeling Framework (SWMF; Tóth et al., 2005) is a high-performance, flexible framework for integrating and coupling multiple physics models and simulations. Developed by the Center for Space Environment Modeling (CSEM) at the University of Michigan and collaborators, the SWMF is primarily used in space physics and space weather, but it is also applicable to a broad range of domains, from solar to atmospheric physics. One of its core components is the Block-Adaptive-Tree Solar-wind Roe Upwind Scheme (BATS-R-US; Gombosi et al., 2021; Powell et al., 1999; Tóth et al., 2012), a versatile and high-performance MHD code with adaptive mesh refinement, which is available through the Community Coordinated Modeling Center (CCMC) at NASA's Goddard Space Flight Center (GSFC). It can be configured to solve the governing equations of not only ideal MHD but also more advanced or generalized formulations, enabling applications across a wide range of plasma environments, while maintaining computational efficiency within the SWMF framework.

In this study, we employed the GM–IE–IM coupled configuration of the SWMF. The global magnetosphere (GM) is modeled using BATS-R-US, while the ionosphere (IE) is described by the Ridley Ionosphere Model (RIM; Ridley et al., 2004), which serves as a standard SWMF ionospheric electrodynamics solver. The inner magnetosphere (IM) is represented by the Rice Convection Model (RCM; Toffoletto et al., 2003). It should be noted that the role of the RCM in this study is to provide a self-consistent inner magnetospheric background rather than to examine detailed ring current dynamics. These components are bidirectionally coupled with time-dependent coupling intervals of 5–10 s. Adaptive mesh refinement is applied, achieving a maximum spatial resolution of 1/8 $R_E$ in the inner magnetosphere. The SWMF simulations are primarily used to investigate the large-scale magnetospheric configuration and FAC structures associated with the TPAs and auroral spirals.

### 3.2. Improved REProduce Plasma Universe (REPPU) Code

The improved REProduce Plasma Universe (REPPU) global MHD simulation code (Nakamizo et al., 2025; Nakamizo and Kubota, 2021) employs a spherical grid system that is unstructured in the angular directions and uses radially increasing grid spacing. This configuration allows for enhanced numerical stability, compared to conventional codes, meaning that the solutions are less prone to divergence due to the absence of distinct singularities. Therefore, such a grid system is expected to facilitate the reproduction of both global-scale transpolar arc (TPA) structures and local-scale auroral spiral structures in a physically consistent manner.

The original REPPU code (Tanaka, 2015) assumes that the Earth's magnetic axis, rotational axis, and the axis normal to the ecliptic plane all agree, that is, they are aligned with the $Z$-axis of the computational domain. In reality, however, the Earth's rotational axis is tilted relative to the ecliptic, and the magnetic axis is further offset from the rotational axis, leading to a complex precessional motion. As a result, direct simulations of the actual geospace environment are not feasible under these simplified assumptions. The improved REPPU code accounts for this precession in an equivalent manner by adjusting the input IMF and solar wind plasma conditions, as well as the solar zenith angle, which affects ionospheric conductance (Nakamizo et al., 2025; Nakamizo and Kubota, 2021).

In the improved REPPU framework, the order of spherical grid subdivision can be freely adjusted depending on system requirements. This flexibility allows the simulations to be performed with higher temporal and spatial resolutions. In this study, we apply 6th-order grid subdivision, resulting in a spatial resolution of approximately 1° in both latitude and longitude. The REPPU simulations take into account the solar wind-magnetosphere-ionosphere coupling and determines ionospheric conductance using proxies such as the FAC intensity and the plasma temperature directly above the ionosphere. Consequently, the ionospheric conductance reflects the simulated magnetotail processes.

## 4. Results

### 4.1. An Overview of Observation

Figure 1 shows a fine example of the auroral spiral (pointed out with a thick yellow arrow) and the TPA contemporaneously observed by the Polar UVI (panels a and b) and the all-sky camera (ASC) installed at the Longyearbyen station (75.32° magnetic latitude and 111.0° magnetic longitude, Figure 1c) at 21:08 UT on 10 January 1997. The long-term movie of the auroral spiral detected by the Longyearbyen ASC is available in the supporting information of Nowada et al. (2023). The auroral spiral clearly identified by both Polar UVI and the Longyearbyen ASC was only one (pointed out with thick orange arrows).

We, hereafter, discuss this auroral spiral identified by satellite- and ground-based observations, and the other auroral signatures azimuthally neighboring on the auroral spiral are referred to as auroral spots (auroral spiral and spots). The auroral spiral was seen to the southwest of Svalbard (Figures 1b and 1c). The nightside end of the dawnside TPA was distorted toward the midnight sector, and the TPA extended from the poleward edge of the postmidnight auroral oval at ∼1h MLT and ∼72° MLat to the prenoon auroral oval across the dawnside of the north pole (Figure 1b; see also the supporting information of Nowada et al., 2023). This type of TPA is referred to as nightside distorted (J-shaped) TPA (see Nowada et al., 2020), but we deal with it simply as TPA in this study.

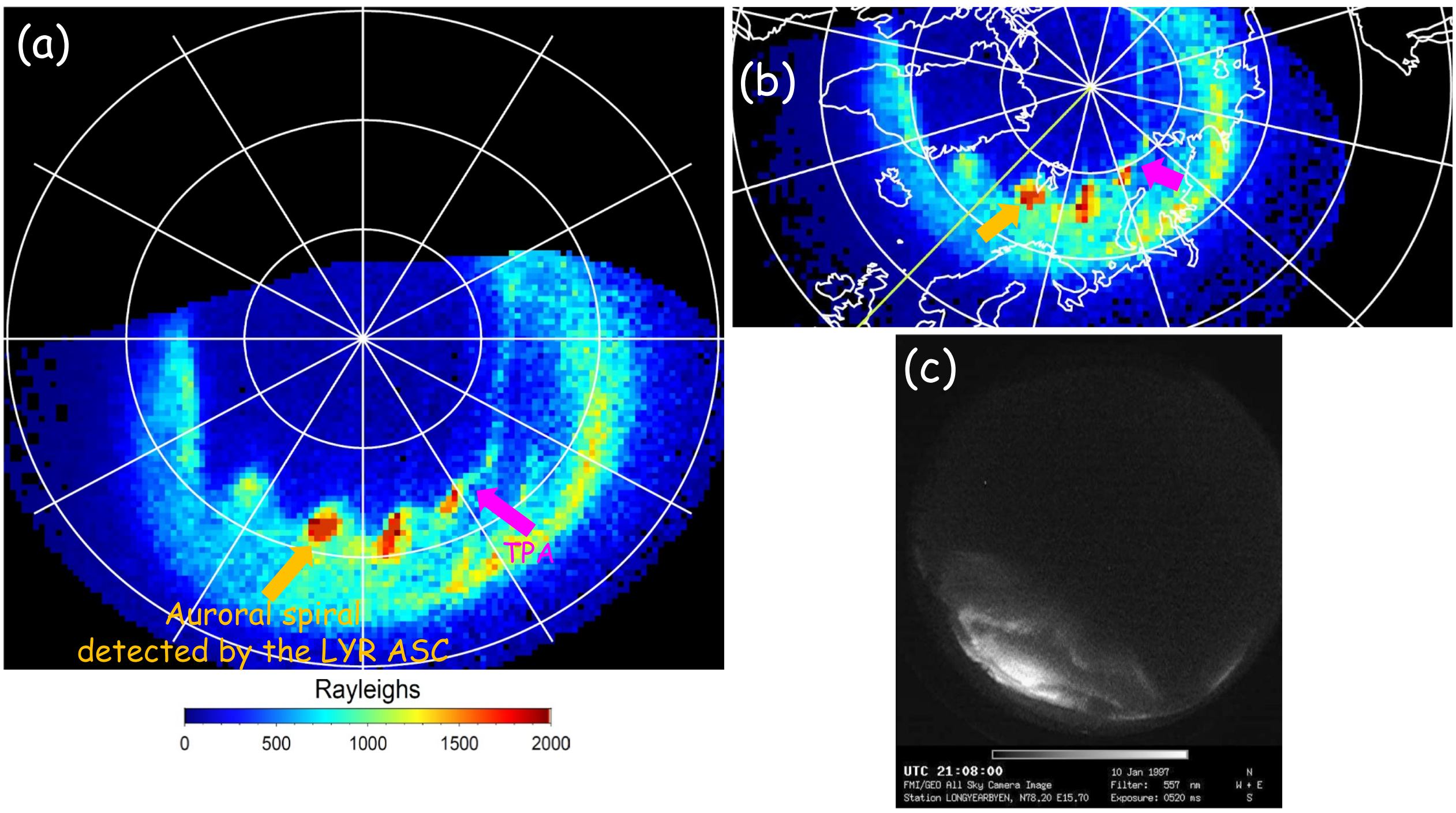

**Figure 1.** Fine snapshots of contemporaneous appearances of the auroral spiral (pointed out by thick orange arrows) and the transpolar arc (indicated with thick magenta arrows) observed by the Polar ultraviolet imager (UVI) and Longyearbyen all-sky camera (ASC) are shown. Panels (a and b) are Lyman-Birge-Hopfield short (LBHS) emission images in altitude adjusted corrected geomagnetic (AACGM; Baker & Wing, 1989) and geographic coordinate systems, respectively, taken with an integration time of 36 s at 21:08:40 UT on 10 January 1997. Panel (a) is oriented such that the bottom, right, top, and left correspond to midnight (0 h magnetic local time; MLT), dawn (6 h MLT), noon (12 h MLT), and dusk (18 h MLT), respectively. The white circles are drawn every 10° from 60° to 80° magnetic latitude (MLat) for panel a and from 50° to 80° geographic latitude for panel (b) The white straight lines are drawn every 2 h in MLT. The color code is assigned according to the auroral brightness in units of Rayleigh. Panel (c) shows an image of the auroral spiral taken from the ASC installed at Longyearbyen for the nearest observational time of panels (a, b).

### 4.2. Solar Wind Conditions

Figure 2 shows the OMNI solar wind parameters and geomagnetic indices for the 6 h interval between 17:00 UT and 23:00 UT on 10 January 1997. From top to bottom, the panels show the SMU and SML indices (Newell and Gjerloev, 2011), the $Y$ and $Z$ components of the interplanetary magnetic field (IMF-$B_Y$ and -$B_Z$) in geocentric solar magnetospheric (GSM) coordinates, the IMF clock angle derived with arctan(IMF-$B_Y$/IMF-$B_Z$), the Akasofu-Perreault parameter ($\varepsilon_{AP}$, a measure of the solar wind energy input rate; Perreault and Akasofu, 1978), and the solar wind dynamic pressure ($P_d$), velocity ($V_{SW}$) in GSM coordinates, and proton number density ($N_P$). The values of $K_p$ are shown at the bottom of the figure. The interval of the contemporaneous appearances of the TPA and the auroral spiral was identified by visual inspection, based on the images from the Polar UVI and the all-sky camera (ASC) at the Longyearbyen station. Again, note that auroral spiral denotes only one that was identified by both satellite- and ground-based observations. The focused time interval (19:59 UT–21:33 UT) is bracketed with a pair of magenta broken lines.

During the presented interval, the whole cycle of a substorm from onset to recovery phases, which was accompanied by moderate geomagnetic disturbances with a $K_p$ range from 3+ to 4, can be seen. Namely, SML shows a sharp, large decrease from ∼18:50 UT (the substorm onset) to ∼19:05 UT, and then SML recovered to ∼0 nT at ∼22:15 UT. Small negative bay variations were seen in SML during the time intervals when the TPA and the auroral spiral were contemporaneously observed (hereafter, referred to as TPA + AS). On the contrast, SMU was almost constant during the presented time interval. Note that TPA + AS were observed even when no significant perturbations of SML and SMU occurred.

**Figure 2.** Plots of solar wind parameters and geomagnetic activity indices during the 6 h interval from 17:00 UT to 23:00 UT on 10 January 1997 are shown. From top to bottom: the *SMU* and *SML* indices which were derived, based on high-latitude geomagnetic field data obtained from the SuperMAG geomagnetic observatory network (Gjerloev, 2012; Newell and Gjerloev, 2011); the IMF-$B_Y$ and -$B_Z$ components in geocentric solar magnetospheric (GSM) coordinates; the IMF clock angle (arctan [IMF-$B_Y$/IMF-$B_Z$]); the Akasofu-Pelleaut parameter ($\varepsilon_{AP}$) computed with $V_{SW}B_T^2\sin^4(\theta_{CLOCK}/2)\,(4\pi L_0^2/\mu_0)$, where $V_{SW}$ is the solar wind velocity in GSM coordinates, $B_T$ is the magnetic field intensity as calculated by sqrt(IMF-$B_X^2$ + IMF-$B_Y^2$ + IMF-$B_Z^2$), and $L_0 = 7.0\ R_E$; the solar wind dynamic pressure ($P_d$); the solar wind speed ($V_{SW}$), and the proton number density ($N_P$). The $K_p$ index is indicated at the bottom of the figure. The interval of contemporaneous appearances of the TPA and the auroral spiral (19:59 UT–21:33 UT) is bracketed with magenta vertical broken lines.

During the TPA + AS interval, clear jumps were seen in the IMF-$B_Y$ and -$B_Z$ components, and the IMF clock angle at 21:10 UT. In particular, the IMF-$B_Z$ component, which remained weakly southward for at least 4 h, showed a sharp polarity change from southward (negative) to northward (positive), while IMF-$B_Y$ had a dominantly dawnward (negative) component, that is, the sign changes were not seen. Due to this abrupt IMF-$B_Z$ jump, the IMF clock angle increased from −90° to −45°. The ε parameter shows a significant decrease, associated with the clock angle's jump, which was preceded by a gradual decrease after 19:59 UT. The associated solar wind dynamic pressure and plasma number density gradually increased by a factor of 2.6, from approximately 2.5 to 6.5 nPa and from about 9.0 to 24.0/cc, respectively, despite only minor variations in the solar wind. Only the solar wind velocity showed no significant variations associated with the auroral spiral.

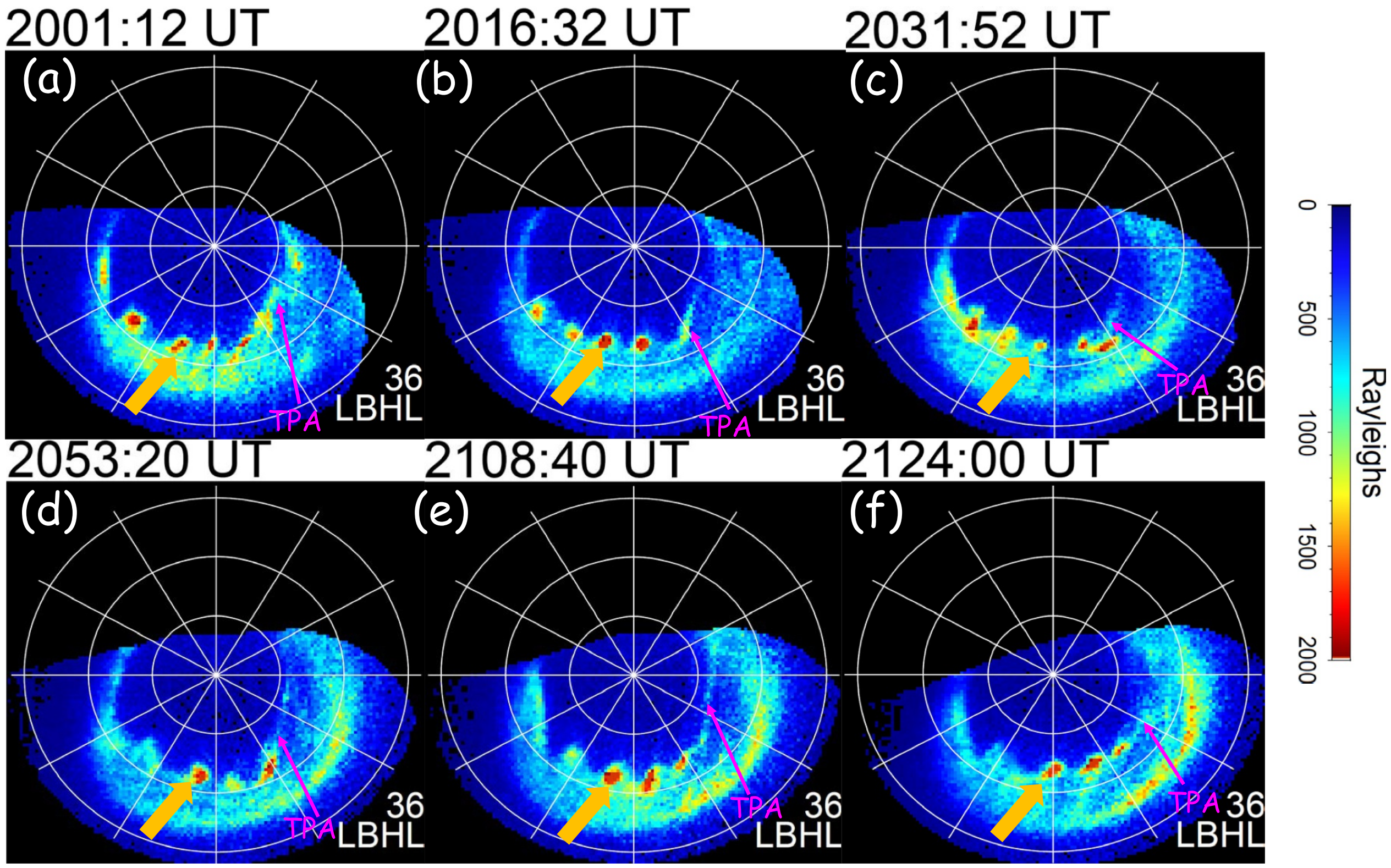

**Figure 3.** Representative snapshots of LBHL 36 s images of the contemporaneous appearances of the TPA and the auroral spiral that was identified with the ASC at Longyearbyen are shown in panels (a–f) at 20:01:12, 20:16:32, 20:31:52, 20:53:20, 21:08:40, and 21:24:00, respectively. The color code is assigned according to the auroral brightness in units of Rayleigh. The white circles are drawn every 10° from 60° to 80° magnetic latitude (MLat), and the white straight lines are drawn every 2 h in MLT. The TPA and the auroral spiral identified are indicated with thin magenta and thick orange arrows, respectively.

### 4.3. Polar UVI Auroral Observations in the Polar Cap

Figure 3 shows the six observation examples of contemporaneous appearance of TPA + AS that were detected by the Polar UVI. All panels in Figure 3 are long Lyman-Birge-Hopfield emission (LBHL; ∼170 nm) images from the Polar UVI with an integration time of 36 s in the same format as Figure 1a. Polar detected the TPA growing from the poleward edge of the nightside auroral oval near 3h MLT to the dayside, which was also aligned with the dawnside auroral oval. This TPA was stable and had no significant dawn-dusk motion within the polar cap. Such a static TPA near the dawn (or dusk) auroral oval, so-called oval-aligned arc, had already been reported and its fundamental characteristics, such as the relation to the IMF orientations, was discussed by Murphree and Cogger (1981) and Kullen et al. (2002, 2015). In addition to the TPA, azimuthally-chained three or four auroral spots, including the auroral spiral identified by the Longyearbyen ASC (Nowada et al., 2023), were seen at the poleward edge of the nightside auroral oval.

### 4.4. Magnetotail Source Regions of Auroral Spiral and TPA

To investigate the source regions of the auroral spiral and the TPA in the nightside magnetosphere, we mapped each pixel data of the Polar UVI images onto the magnetic equatorial plane, based on the Tsyganenko 96 empirical magnetic field model (T96; Tsyganenko and Stern, 1996). The technical details on mapping are described in Nowada et al. (2023). Figure 4 shows the projections of the Polar UVI images of the TPA (highlighted with a magenta oval) and the auroral spiral (surrounded with an orange oval) onto the GSM-X-Y nightside magnetic equatorial plane. Note that the nightside equatorial plane mapping of the auroral images sensitively depends on magnetic field models (Lu et al., 2000).

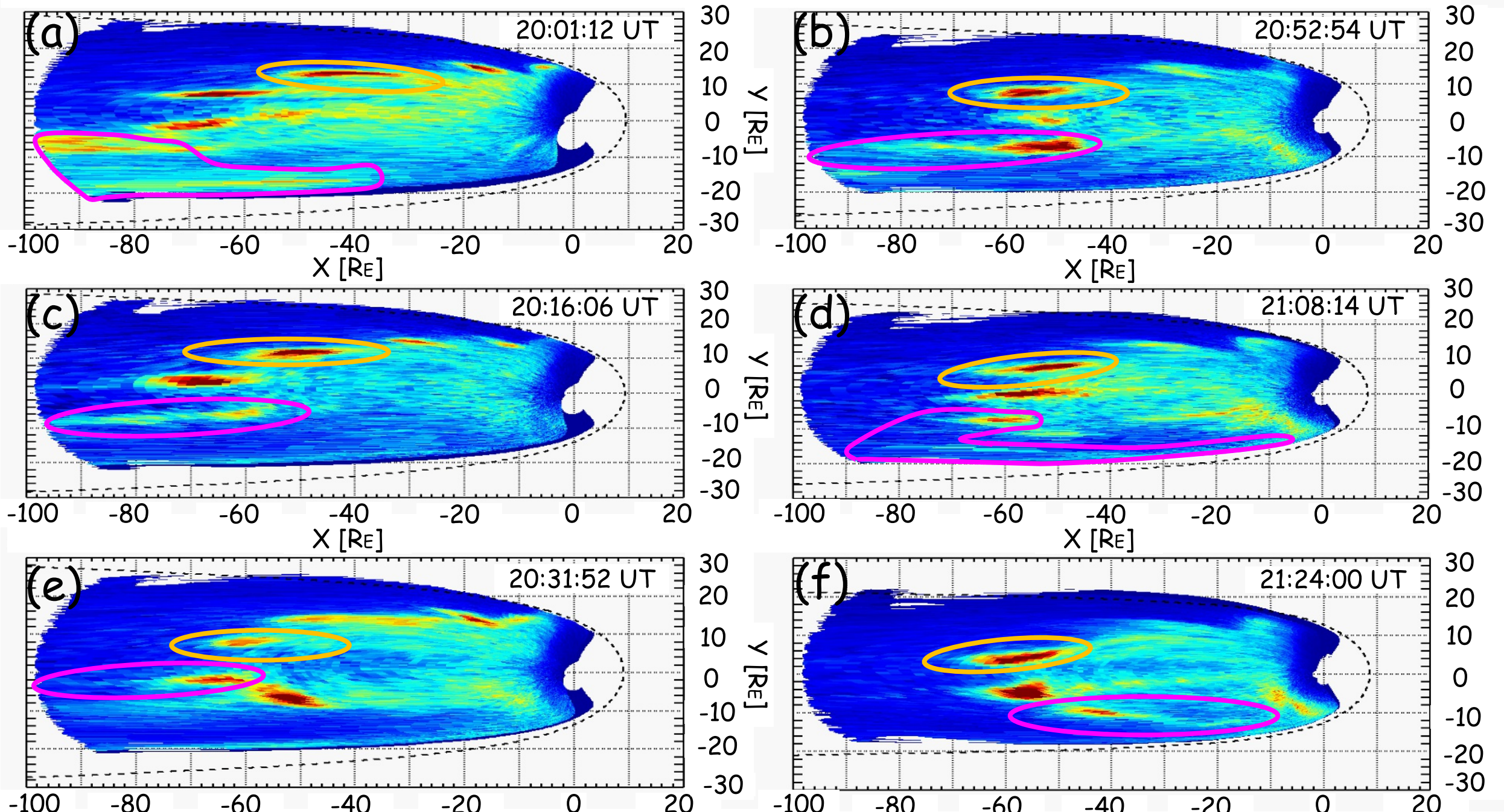

**Figure 4.** Six representative magnetotail magnetic equatorial projections of Polar UVI auroral pixel data for the contemporaneous appearances of the TPA and the auroral spiral are shown in panels (a–f) at 20:01:12, 20:16:06, 20:31:52, 20:52:54, 21:08:14, and 21:24:00, respectively. Each UVI pixel was traced to the magnetic equatorial plane, based on the Tsyganenko 96 geomagnetic field model (Tsyganenko and Stern, 1996), using the calculation routines implemented in the Space Physics Environment Data Analysis Software (SPEDAS) 4.1 package (Angelopoulos et al., 2019). The color code is assigned according to the auroral brightness in units of Rayleigh, and its range is the same as Figures 1 and 3. Magenta and orange ovals highlight the magnetotail source regions of the TPA and the auroral spiral identified with the Longyearbyen ASC, respectively. The dashed black curves indicate the model magnetopause locations derived from Shue et al. (1998). The T96-based mapping is shown as a qualitative reference, and the physical interpretation in this study is primarily based on the results of the self-consistent MHD simulations.

Based on the mapping onto the magnetic equatorial plane, the source regions of the TPA exhibited a tailward-elongated structure, with scales ranging approximately 40–85 $R_E$ in $X$ and from 6 to 16 $R_E$ in Y (in GSM coordinates), as indicated by magenta ovals. It is interesting that the TPA source region lay on the dawnside magnetotail equatorial plane. In particular, in Figures 4a and 4d, the TPA source region extends tailward about $60 \sim 80$ $R_E$ and also toward the midnight meridian. Note that this projection of the TPA using the T96 field line model may differ from the actual magnetic mapping, because the model does not describe well magnetotail deformation and the associated field line twisting due to the intense IMF-$B_Y$ effect that are necessary for TPA formation (e.g., Nowada et al., 2020). The mapping based on the T96 field line model should be regarded as a qualitative reference for visualizing the (large-scale) correspondence between the observed auroral features and the magnetotail configuration. The physical interpretation in this study does not rely on the T96 mapping itself, but instead on the results of the self-consistent global MHD simulations discussed in the following sections.

The projection of the auroral spiral also exhibited a tailward elongation, with scales of ~30 $R_E$ and ~6 $R_E$ in the $X$ and Y directions in GSM coordinates, respectively, even though the spiral had the form of a spot in the ionosphere. This elliptic form is consistent with the discussions by Kaufmann et al. (1990) for quiet time magnetosphere and by Lu et al. (2000) for a substorm. The auroral spiral had almost the same scale as that seen at different times on the same day (Nowada et al., 2023).

To discuss the detailed formation process of these auroral phenomena, we also performed global magnetohydrodynamic (MHD) computer simulations using the SWMF with BATS-R-US. Figure 5 shows the distributions of the field-aligned current (FAC) at the magnetic equatorial plane of the magnetotail. Each time label is almost the same as that in Figure 4. In most panels in Figure 5, velocity reversal from tailward to earthward was observed between $X = -25$ and $-30\ R_E$ near midnight, suggesting the occurrence of magnetotail reconnection and also implying that the generation of flow shear that may drive the TPA-associated FACs. Downward (directed into the plane) FAC structures elongated tailward were clearly reproduced in the dawn sector (Figures 5b and 5d–5f) and near the midnight meridian (Figures 5a and 5c).

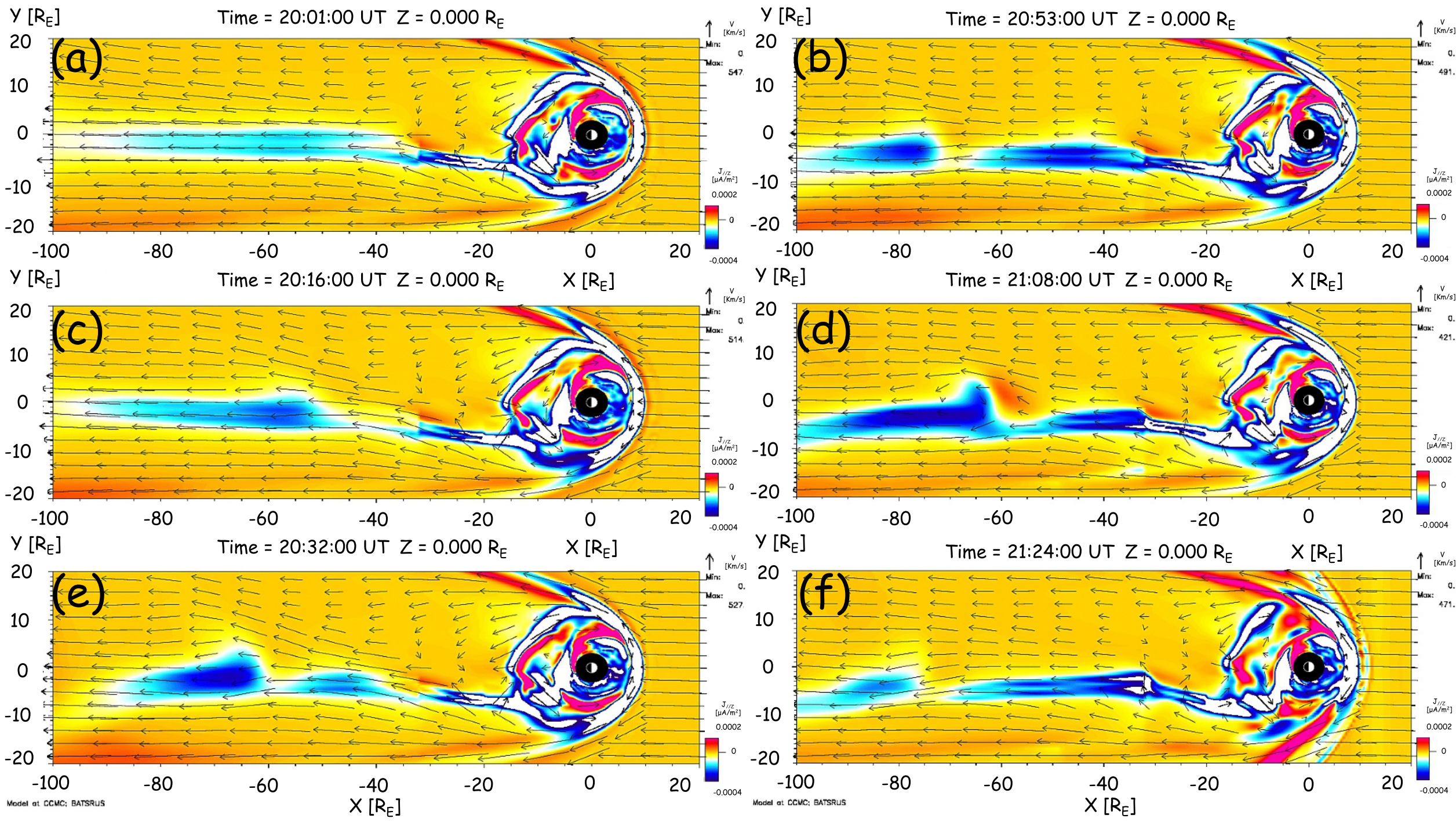

**Figure 5.** The density of field-aligned current (FAC) on the magnetic equatorial plane over the ranges of $X = 20$ to $-100\ R_E$ and $Y = 20$ to $-20\ R_E$ are shown in panels (a–f) at 20:01:00, 20:16:00, 20:32:00, 20:53:00, 21:08:00, and 21:24:00, respectively. The FAC structures are reproduced by Space Weather Modeling Framework with Block Adaptive Tree-Solar wind-Roe Upwind Scheme (BATS-R-US) code provided by the Community Coordinated Modeling Center (CCMC). The color code is assigned according to the Z component (the vertical component perpendicular to the X–Y plane, defined as positive upward and negative downward) of the FAC ($J_{//Z}$) on the magnetic equatorial plane in units of μA/m$^2$. The vectors indicate the plasma flow velocity ($V_{XY}$) projected onto the magnetic equatorial plane in units of km/s. The times of the presented FAC maps are the same as those in Figure 4.

The intensity of these TPA-associated FACs ($J_{//Z}$) was between $\sim 1.5 \times 10^{-4}$ and $\sim 4.0 \times 10^{-4}$ μA/m$^2$. These FAC sheets plausibly correspond to the projections of the TPA shown in Figure 4. The tailward-elongated FAC structures in Figure 5 represent the broad spatial FACs associated with the TPA. As shown in (Figure S1 of Supporting Information S1), for this event, the structured FACs on closed field lines, representing a likely primary FAC source, may extend tailward to $X = \sim -50\ R_E$. The more distant tailward FAC extension likely corresponds to currents flowing along severely stretched and distorted magnetic field lines in the downtail, consistent with the overall dawn-dusk asymmetric configuration. Namely, the observational cases shown in Figures 4b, 4c, and 4e seem to be explained well by the MHD simulations (Figures 5b, 5c, and 5e). On the other hand, the FAC sheet profile of the TPA in the magnetotail reproduced by the MHD simulations (Figures 5a and 5d) is possibly different from that of the Polar observation (Figures 4a and 4d), because, for the Polar observation, each UVI pixel data was simply traced to the magnetic equator along the T96 model field line without considering magnetotail twisting due to significant IMF-$B_Y$ effects. According to a TPA formation model derived by magnetotail observations (Nowada et al., 2020; particularly see their Figure 7), as the TPA is growing to the dayside polar cap, the corresponding TPA-associated FAC structure should extend tailward, deformed by the IMF-$B_Y$ component. Hence, the simulation result can explain well the TPA formation model proposed by Nowada et al. (2020).

The ion velocity vectors show that no significant (anti-)clockwise vortices which should be the source of the FACs from the magnetotail to the ionosphere are absent near/around possible auroral spiral source regions identified by the Polar UVI observations. Unfortunately, these whole magnetotail MHD views do not show significant tailward elongating structures that correspond to the auroral spiral and spots as detected by Polar. Furthermore, even taking look at the ionospheric FAC distributions that are simply projected the calculated FAC results in the magnetotail onto the ionosphere along the model field line, no clear azimuthally-chained auroral spiral and spots could be found at the poleward edge of the nightside auroral oval (not shown).

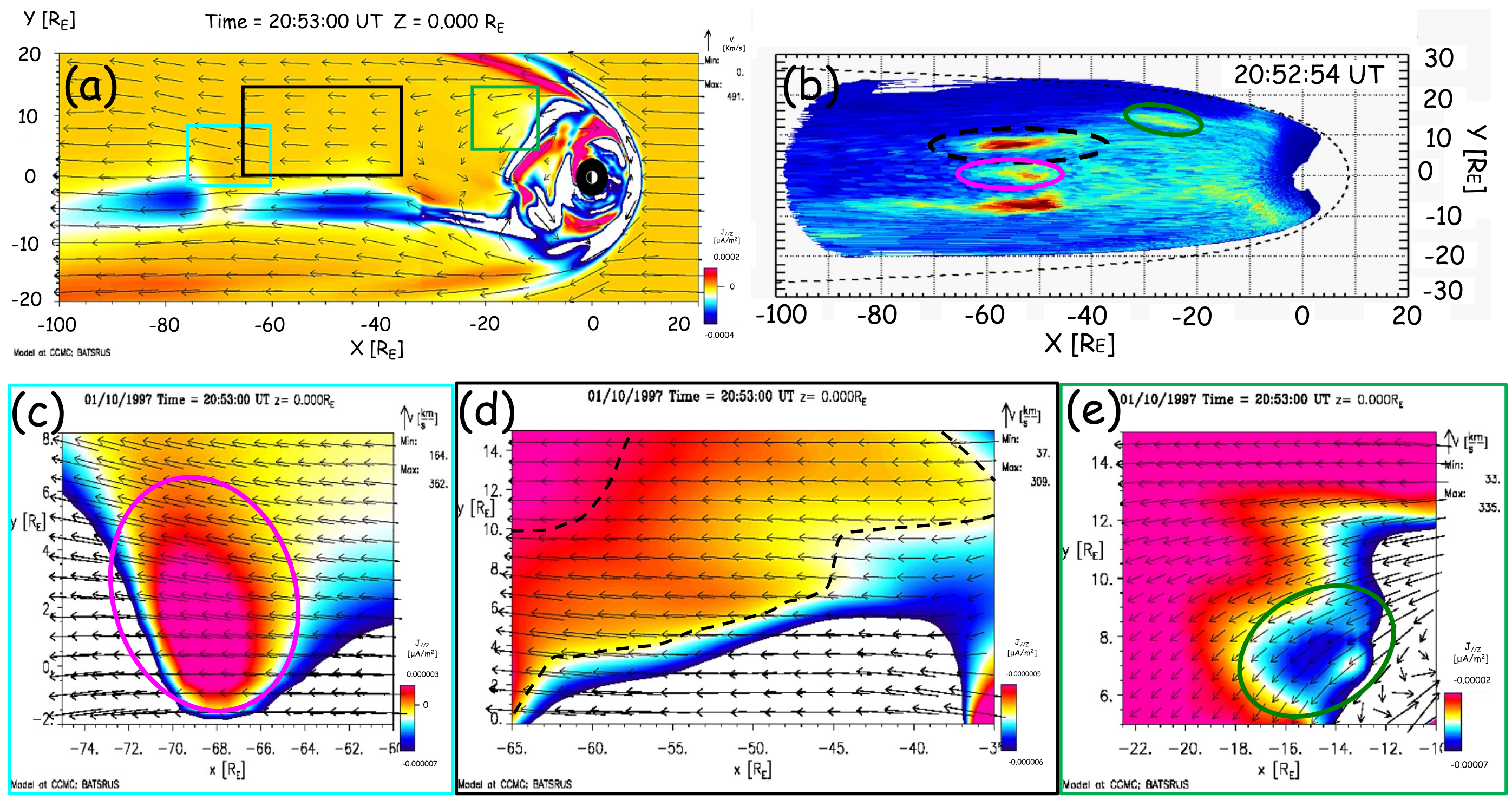

**Figure 6.** An example of comparison between global MHD (panel a) and observational (20:52:54 UT, panel b) results, and zoomed-in plots of FAC sheets reproduced by Space Weather Modeling Framework with BATS-R-US global MHD simulations (20:53:00 UT, panels c–e) are shown. The color codes in panels (a) and (c–e) are assigned according to $J_{//Z}$ in units of $\mu A/m^2$, and that in panel (b) is assigned according to the auroral brightness in units of Rayleigh with the same range as Figures 1, 3, and 4. The color code scales of panels (c–e) are different from that in panel (a). Also note that the color codes in panels (d) and (e) indicate the intensity of the downward FACs. The cyan, black, and green squares in panel (a) correspond to the regions shown in panels (c–e). The regions surrounded with cyan, broken black, and green curves in panels (c–e) correspond to the regions highlighted with the ovals in the same colors in panel (b).

A comparison of the results between the Polar UVI observation and the SWMF with BATS-R-US MHD simulations, and zoomed-in plots of the MHD simulations on 20:53 UT are shown in Figure 6. The tailward-elongated TPA source region was observed in Figure 6b, and its FAC sense was from the ionosphere to the magnetotail (into the plane), which was reproduced by global MHD simulations as shown in Figure 6a. The Y location (about $-8 \sim -6\ R_{\mathrm{E}}$) where the source region was lying was also consistent with the simulations. On the other hand, the spiral source region seen by the Polar observation (Figure 6b) was not seen in the simulations (Figure 6a).

We then examined in detail the MHD simulation results by changing the FAC scale and zooming in possible spiral source regions. Figures 6c–6e show the zoomed-in plots of the MHD simulation results in the regions marked with cyan, black, and green squares in Figure 6a. The color scales are different from Figure 6a and each of three panels. In all three panels, the velocity vectors show no significant flow shears (vector difference), including flow vortices that can drive FACs, were not seen. Instead, a very small upward FAC structure (Figure 6c) and downward FAC structures (Figures 6d and 6e) were detected in each region. Their intensities were $\sim 3.0 \times 10^{-6}$ (Figure 6c), $\sim 5.0 \times 10^{-7}$ (Figure 6d), and $\sim 7.0 \times 10^{-4}\ \mu A/m^2$ (Figure 6e). They tended to decrease with increasing distance down the magnetotail. The auroral spiral and its adjacent auroral spots, surrounded by the magenta, broken black, and green ovals in Figure 6b, might correspond to the FAC structures marked in the same colors in Figures 6c–6e. In particular, the region in Figure 6d should correspond to the spiral source region and is elongated tailward, compared with the other two regions. Given this simulation result, the FAC associated with the spiral was much smaller than the background and TPA-associated FACs. The intensity of the TPA-associated FACs is approximately three orders of magnitude greater than that of the FACs associated with the auroral spiral. The results of comparison between the Polar observation and MHD reproductions on 20:16 UT and 21:08 UT as shown in Figure 5 can be seen in Figures S2 and S4 of Supporting Information S1. To form these very weak spiral-associated FACs near $X = \sim -70\ R_E$, a flow shear between high-speed plasma flows generated by magnetotail reconnection and background slower plasma is likely required. As shown by the meridional magnetic field vector plots overlaid on the FAC distributions in Figure S6 of Supporting Information S1, taken along a cut at $Y = 7.0\ R_E$ corresponding to the location of the auroral spiral shown in Figure 6d, rotational discontinuities are identified along the plasma sheet/current sheet, which is consistent with the magnetic field topology expected during magnetic reconnection. The trains of rotational discontinuities persist down to $X = \sim -120\ R_E$, suggesting that the reconnection site is located tailward of this region. A possible reconnection region, identified by the field line reversal, is highlighted by the magenta square. This distant magnetotail reconnection may serve as a source of the spiral-associated FACs seen near $X = -70\ R_E$, suggesting that the closed magnetic field line boundary extends at least to $X = \sim -120\ R_E$.

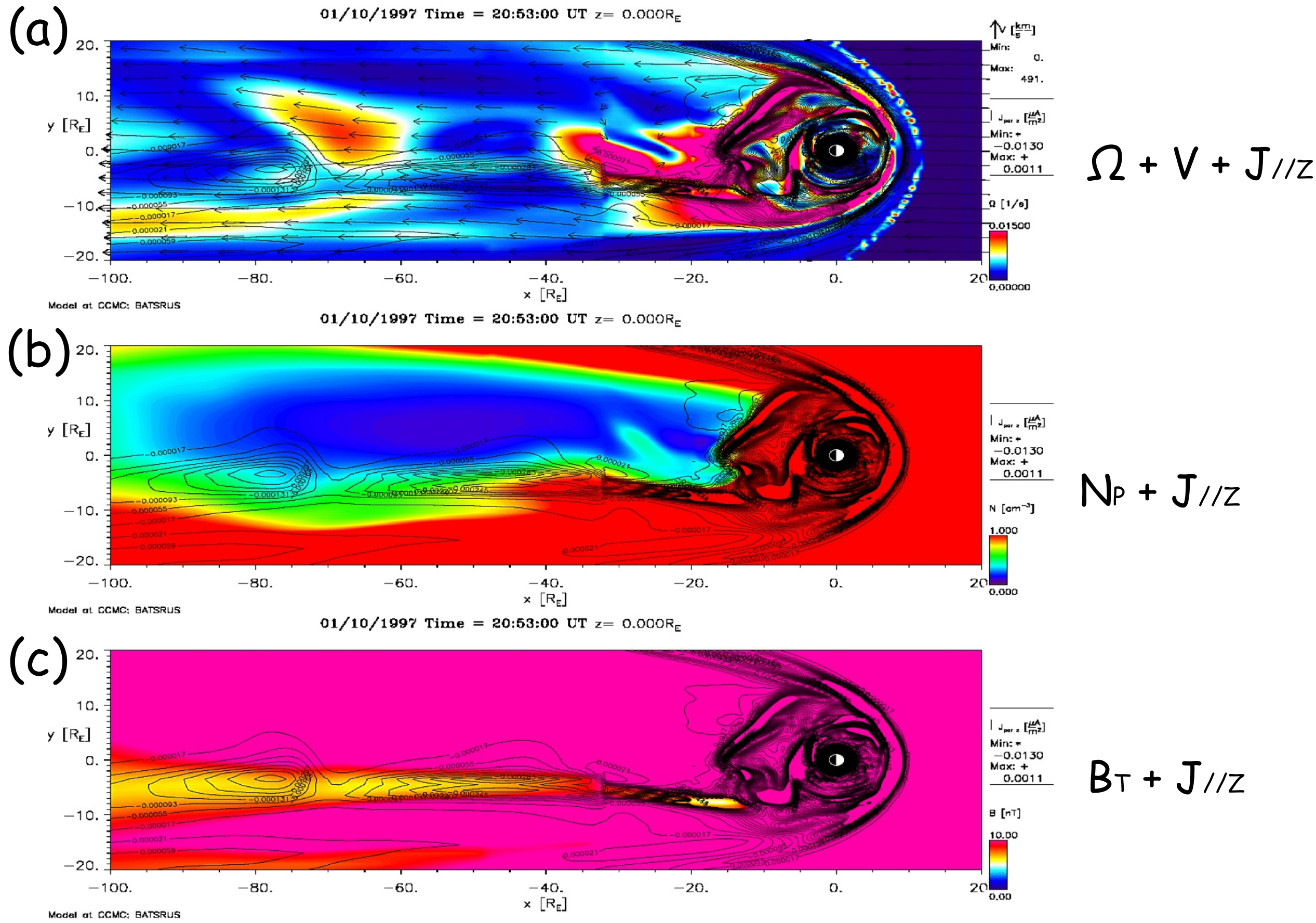

**Figure 7.** Vorticity (panel a), plasma density (panel b), and magnetic intensity (panel c) maps on the magnetotail equatorial plane, which are reproduced by Space Weather Modeling Framework with BATS-R-US MHD simulations, are shown. The plotted time is the same as that in the MHD simulation results in Figure 6. The contours in each panel show $J_{//Z}$. In panel a, velocity vectors are also superimposed.

To address why the spiral-associated FAC structures are so small, we investigated the distributions of the vorticity (Figure 7a), the plasma (ion) density (Figure 7b), and the magnetic field intensity (Figure 7c) on the magnetic equatorial plane on the same time as Figure 6. In the region shown in Figure 6c, intense vorticity (indicated with reddish and yellow colors) can be seen while in the regions of Figures 6d and 6e, much weaker vorticity was predominant as indicated with bluish colors. In the dawnside magnetotail, clear vortices were observed in the TPA-associated $J_{//Z}$ region extending from $\sim -10$ to $-40\ R_E$. In the TPA source region further down the tail beyond $-40\ R_E$, however, these vorticities weakened significantly.

Taking a look at the plasma density distribution on the magnetotail equatorial plane, strong dawn-dusk asymmetry of the density profile was found. The TPA source region primarily lay from middle (greenish color) to high (reddish color) plasma density region, while the low plasma density was dominantly distributed over the auroral spiral source region located on the duskside, indicated with bluish color. The region of low magnetic field intensity (yellowish color) was spatially aligned with the TPA source region and was located where the plasma sheet or neutral sheet is typically found in the nightside magnetosphere. The regions corresponding to the auroral spiral sources (on the duskside) had an intense magnetic field strength. The profiles of these physical parameters should strongly affect the FAC structures associated with the TPA and the auroral spiral. The results on 20:16 UT and 21:08 UT are shown in Figures S3 and S5 of Supporting Information S1.

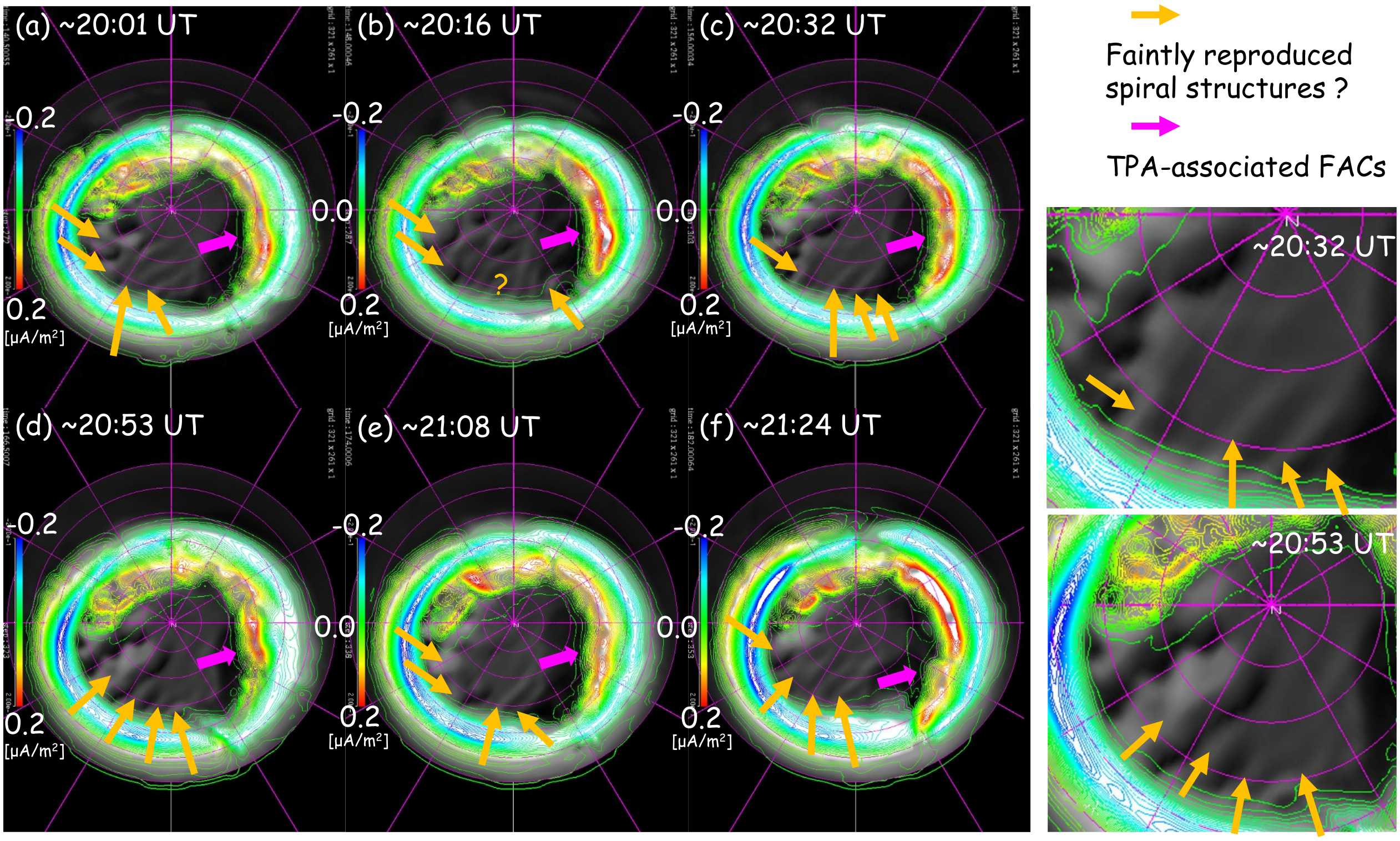

**Figure 8.** Representative zoomed-in snapshots of aurora-associated FAC distributions and ionospheric conductance during the interval of contemporaneous appearances of the TPA and the auroral spiral, reproduced by the global MHD simulations using the improved REPPU code are shown. Color contours represent the ionospheric FACs, while the grayscale (gray-to-white) shading indicates the ionospheric conductance determined by the corresponding magnetotail processes, serving as a proxy for auroral structures. Reddish (bluish) contours indicate FACs flowing toward (away from) the ionosphere. Magenta circles are drawn every 5° from 65° to 90° magnetic latitudes, and magenta radial straight lines indicate magnetic local time at 2 h intervals. Faint streak-like structures, interpreted as possible auroral spiral and auroral spots reproduced by the improved REPPU simulations, are indicated by thick orange arrows, while the TPA-associated FACs are marked by magenta arrows.

Figure 8 shows representative examples of overlaid plots of the FAC distributions (colored contours) and the ionospheric conductance, which is determined by the corresponding magnetotail processes reproduced by the improved REPPU code and shown with grayscale (gray-to-white). The grayscale patterns can be interpreted as auroral structures, since enhanced conductance reflects increased particle precipitation in the ionosphere. Brighter (whiter) shading therefore indicates regions of higher conductance and potentially more intense auroral precipitation, but does not necessarily correspond to intense FACs, which are represented separately by the colored contours. During the time interval presented, several faint but continuous streak-like structures (indicated by orange arrows) were identified within the polar cap, apparently extending from the poleward edge of the nightside auroral oval toward the dayside. Ambiguous structures, marked by an orange question mark, were also seen, for which it remains unclear whether they form coherent streak-like structures. Nevertheless, these faint structures may correspond to simulated auroral spiral and spots, although their morphology does not appear as distinctly "spot-like" (see zoomed-in plots at ∼20:32 UT and ∼20:53 UT). Clear FAC structures associated with the spiral (and auroral spots), as indicated by the colored contours, however, were not reproduced in the simulations.

Spiral structures might be reproduced more clearly, if the temporal and spatial resolutions of the REPPU simulations were further enhanced. In contrast, the TPA structures are persistently present at the poleward edge of the dawnside auroral oval throughout the interval, and their source regions in the magnetotail can be successfully identified, as indicated by the thick magenta arrows in Figure 8; Figure S7 in Supporting Information S1. The ionospheric TPA-associated FAC system is, however, relatively weak and appears adjacent to the dawnside auroral oval near its poleward edge, exhibiting no clearly defined current polarity. Nevertheless, the ionospheric and magnetospheric FACs exhibit consistent polarity and spatial correspondence, indicating that the TPA-associated FAC system is present but remains weakly developed. In particular, the spatial distribution of the FAC system is broadly consistent between the SWMF with BATS-R-US and the improved REPPU simulations, despite differences in the detailed structures between the two models. In contrast, the spiral-associated FAC source regions in the magnetotail are not well reproduced in this model (Figure S7 in Supporting Information S1). These results suggest that the TPA is accompanied by a global-scale FAC system in both the magnetosphere and the ionosphere, whereas the magnetotail FAC structures associated with the auroral spiral are likely weaker in intensity. Although the spiral-associated FACs appear to elongate tailward in the observations, they remain spatially localized in the ionosphere. Such characteristics are consistent with the magnetotail spiral-associated FAC structures reproduced by the SWMF with BATS-R-US simulations (Figure 6; Figures S2 and S4 in Supporting Information S1). This consistency supports the interpretation that the spiral-associated FAC system is present but intrinsically weaker and less organized than the TPA-associated FAC system.

The ionospheric FAC distributions ($J_{//}$) at the same times as those shown in Figure 5 are also reproduced, based on the SWMF with BATS-R-US (see Figure S8 in Supporting Information S1). In this case, neither the TPA nor the auroral spiral (and auroral spots) observed by Polar UVI (Figures 1 and 2) is reproduced. The FAC systems appearing in the duskside exhibit clear downward currents, indicated by reddish colors, inconsistent with the TPA-associated FAC polarity inferred from the magnetotail FACs in the SWMF with BATS-R-US (Figure 5). The spiral (or spot-like auroral structures) is also absent at the poleward edge of the nightside auroral oval. For the TPA in particular, it is well-known that global MHD models (except for the improved REPPU) often fail to reproduce fine-scale TPA-associated FAC structures in the ionosphere. This limitation also noted in previous TPA simulation studies (e.g., Hill et al., 2024; Kullen and Janhunen, 2004; Wang et al., 2025). Therefore, the ionospheric TPA-associated FAC structures could not be reproduced by the SWMF with BATS-R-US model.

### 4.5. Ground-Based Observations Associated With the Auroral Spiral

We investigated and argued the case of contemporaneous appearances of the TPA and the auroral spiral, based on the Polar UVI observations and the global MHD simulations using the SWMF with BATS-R-US and improved REPPU code. Besides, using equivalent ionospheric currents (EICs) derived from geomagnetic field variations from the International Monitor for Auroral Geomagnetic Effects (IMAGE) ground observatory network (Tanskanen, 2009), we can estimate a profile of the FACs associated with the spiral (and the TPA). Figure 9 shows the EIC vectors every 1 min from 21:05 UT to 21:10 UT. EICs were reconstructed by fitting spherical elementary currents to the measured magnetic field after decomposing the pure ionospheric current contribution and the telluric current component from the raw IMAGE geomagnetic field data. This EIC reconstruction technique was designed by Vanhamäki and Juusola (2020). The plotted EIC vectors roughly correspond to derivations of the ionospheric horizontal magnetic field components (local magnetic northward and eastward components) by 90° clockwise. During the present interval including the time when the spiral and the TPA appeared contemporaneously (particularly, 21:08 UT), counter-clockwise rotations of the EIC pattern is seen at the spiral, indicated by curved green arrows. These EIC profiles enable us to estimate that FACs flew upward from the ionosphere to the magnetotail associated with the spiral (as highlighted by bluish color). These FAC and EIC senses are consistent with that of the FACs in a local MHD simulation by Lysak and Song (1996) and a series of spirals (Davis & Hallinan, 1976; Partamies et al., 2001a). Furthermore, as shown in Figure 6, weak equatorward FAC structures emerged in the spiral-associated source region in the magnetotail, consistent with these spiral FAC profiles estimated from the EIC maps. Because most part of the TPAs observed before, during, and after the spiral's duration lay over the Arctic Ocean, the magnetometers could not cover the entire TPA.

Figure 10 shows deviations of the $B_N$ component (local magnetic northward direction) from all available ground magnetometer observatories near the auroral spiral from 19:45 UT to 21:45 UT, including the time interval of contemporaneous appearances of the TPA and the spiral from 19:59 to 21:33 UT (bracketed with green vertical broken lines). The $B_N$ deviations were derived by subtracting the DC magnetic field component from the observed magnetic field. The plots of deviations of $B_E$ (local magnetic eastward direction) and $B_Z$ (vertical downward direction) at the same observatories during the same time interval are shown in Figure S9 of Supporting Information S1.

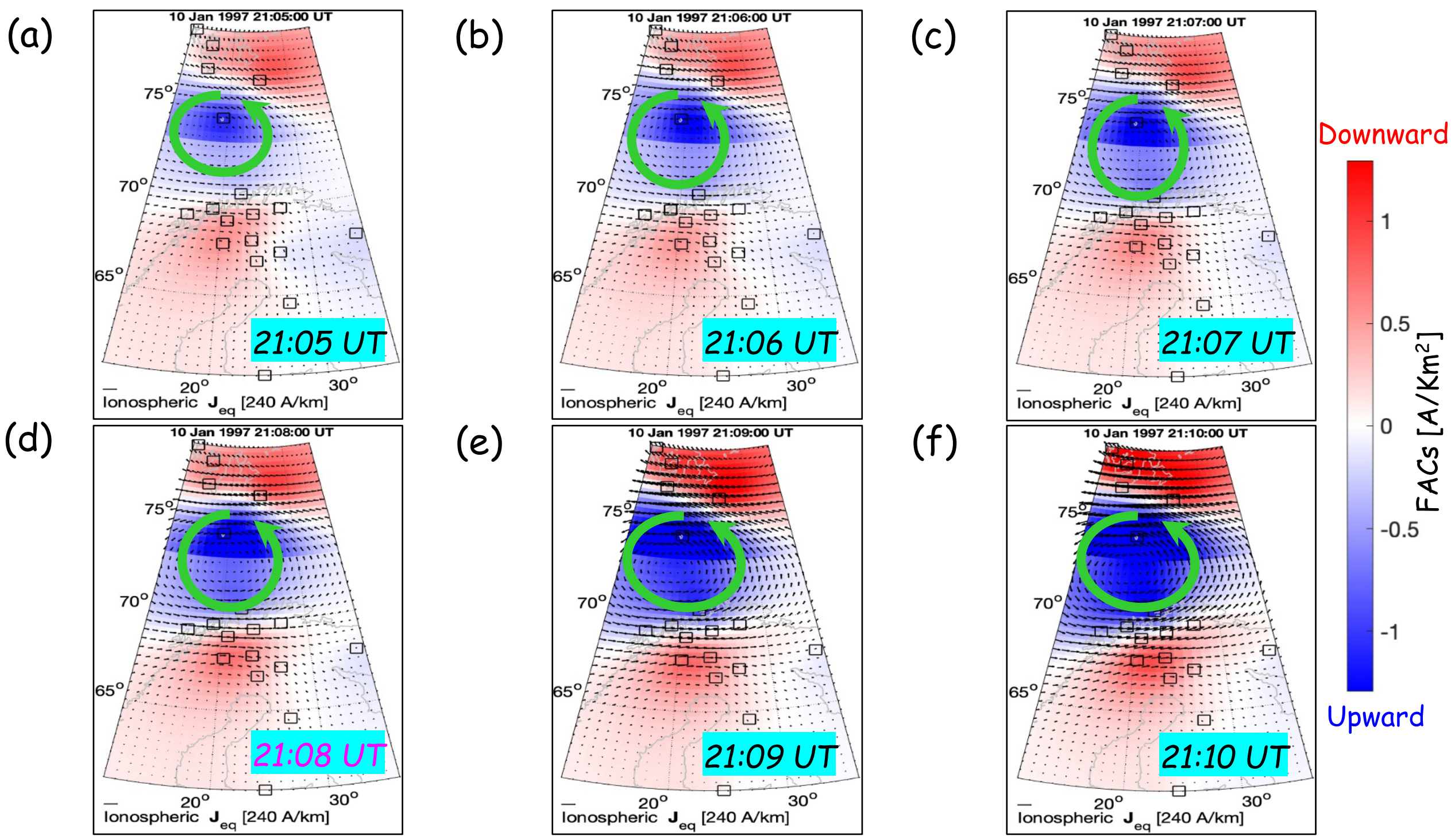

**Figure 9.** Counter-clockwise rotational vortex-like structures of the equivalent ionospheric electric current (EIC) indicated by curved green arrows and derived from the observations of the IMAGE ground observatory network (Tanskanen, 2009) near the auroral spiral region from 21:05 UT to 21:10 UT are shown at a 1 min time step. Panels (a–f) show EIC maps and the associated field-aligned current (FAC) distributions at 21:05, 21:06, 21:07, 21:08, 21:09, and 21:10 UT, respectively. These EIC vectors (black vectors) are derived by extracting the pure ionospheric current contribution by removing the telluric components from the measured geomagnetic field data, based on the technique proposed by Vanhamäki and Juusola (2020). The color code is assigned according to the FAC intensity (in unit of A/km$^2$) and orientation estimated by the EIC vectors.

The $B_N$ component and the other two components at most observatories located in the northern part of the Scandinavia Peninsula seem to show the wavy perturbations in the ultra-low-frequency (ULF) range during contemporaneous appearances of the TPA and the auroral spiral, rather than the deviations seen in the northern (cyan) and western (orange) parts of the spiral. This result implies that the ULF waves as seen equatorward of the spiral, corresponding to the near-Earth magnetotail, might play a role in the spiral formation (appearance). To identify whether or not the observed wavy perturbations are ULF waves, we calculated power spectra of the geomagnetic field deviations at some ground observatories near and equatorward of the auroral spiral.

Figure 11 shows deviations of the $B_N$ and $B_E$ components and their power spectrograms at three ground magnetometer observatories near the auroral spiral (indicated by the three magenta dots in the UVI plot) and equatorward of the spiral (the three yellow dots in the UVI plot) during contemporaneous appearances of the TPA and the spiral from 19:59 UT to 21:33 UT. The power spectrograms were calculated from the wavelet analysis (Torrence and Compo, 1998). The frequency band of Pc5 ultra-low-frequency (ULF: 1.67–6.67 mHz) waves is bracketed by green horizontal broken lines. The plots of deviations of $B_E$ and $B_Z$ and their power spectrograms at the same stations during the same time interval are shown in Figure S10 of Supporting Information S1. At the stations near (around) the auroral spiral, spectrograms of $B_N$ deviations exhibited perturbations in the Pc5 ULF frequency range. No coherent ULF wave packets, however, were clear. Equatorward of the spiral (in the northern part of the Scandinavia Peninsula), deviations of the $B_N$ and ABK-$B_E$ components exhibited the coherent ULF wave packets and the Pc5 band perturbations in their power spectrograms, implying that some ULF wave activity may play a role in generating source particles of the auroral spiral. Note that deviations and their power spectrograms in the $B_Z$ component equatorward of the spiral (see Figure S11 in Supporting Information S1) did not show clear Pc5 wave activity.

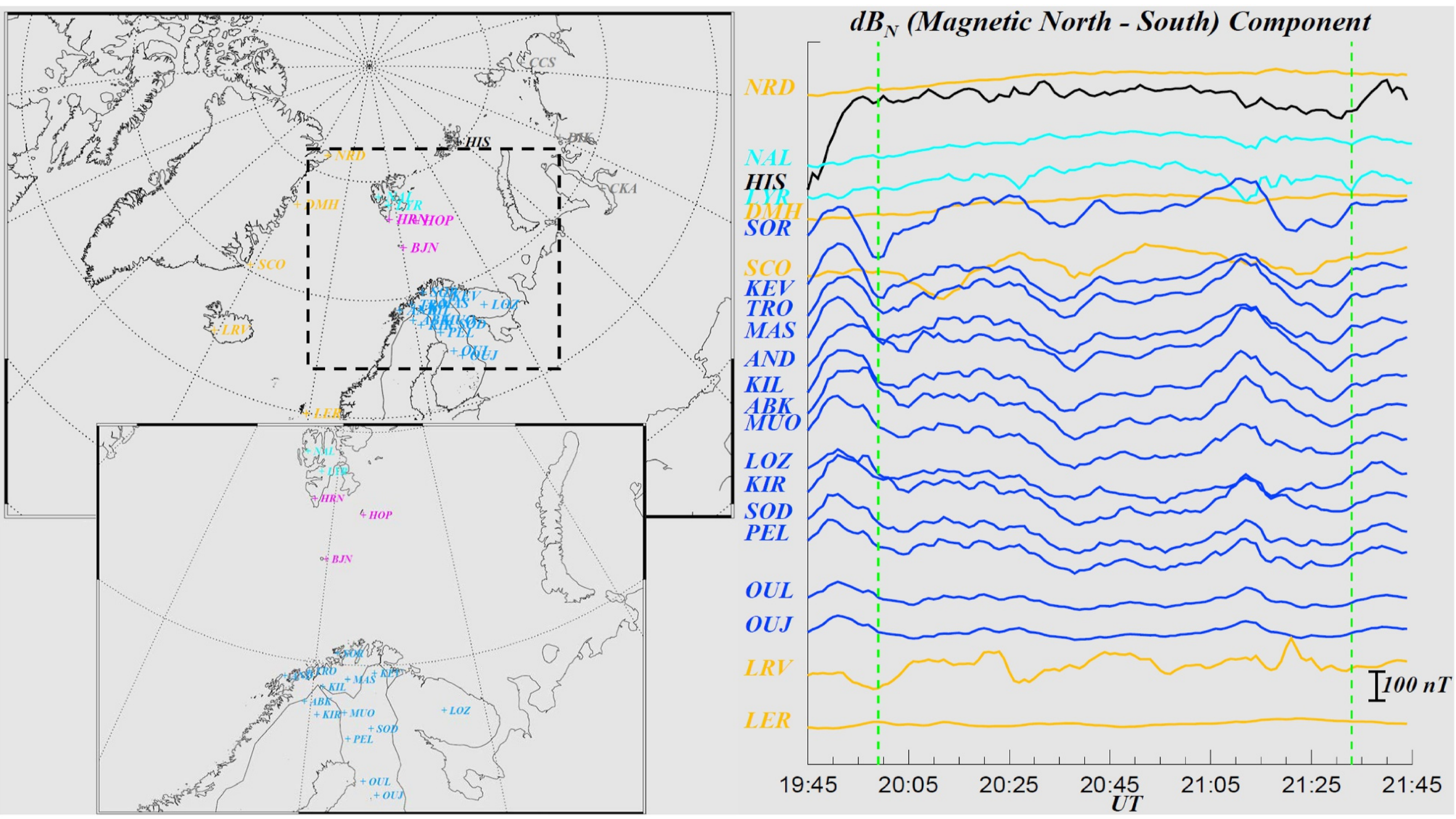

**Figure 10.** Map that shows the locations surrounding the auroral spiral near Svalbard is shown in the left panel. Deviations of the N magnetic field component ($B_N$, local magnetic northward component) at the stations around the Svalbard during the auroral spiral interval from 19:45 UT to 21:45 UT are shown in the right panel. The station codes and plotted magnetic field data for the northern part of the Scandinavia Peninsula, northward (poleward) of the spiral, near the edge of the spiral, westward of the spiral, and eastward of the spiral are shown in blue, cyan, magenta, orange, and black, respectively. The time interval when the TPA and the spiral contemporaneously appeared (from 19:59 to 21:33 UT) is bracketed with two green vertical broken lines.

## 5. Discussion and Conclusions

Intriguing contemporaneous appearances of the global-scale TPA and local-scale auroral spiral were observed during the late stage of the substorm recovery phase and under the intense dawnward IMF-$B_Y$ with the negative-to-positive IMF-$B_Z$ on 10 January 1997. During this interval, no particular solar wind plasma variations was seen. Previously, Nowada et al. (2023) investigated auroral morphological changes after the substorm expansion phase, particularly auroral spiral formation, for the present event and clarified how the auroral spiral and its source in the magnetotail were formed. In this study, we attempted to explain how the magnetotail and ionospheric structures, particularly FAC structures, are when the TPA and spiral with different scales contemporaneously appeared. To examine this, we performed global MHD simulations using two different types of MHD code of SWMF with BATS-R-US and improved REPPU. We then investigated to what extent the Polar UVI results can be addressed with these MHD simulations. Global MHD simulations based on these two MHD codes reproduced the FAC structures associated with the global-scale TPA in the magnetotail (Figure 5; Figure S7 in Supporting Information S1) and in the polar cap (Figure 8). Meridional and cross-sectional investigations of the simulation results (Figure S1 in Supporting Information S1) suggest that the dominant part of the TPA-associated FAC system is concentrated near $X = \sim -50\ R_E$. This refined interpretation is enabled by examining the self-consistent magnetospheric topology under intense IMF-$B_Y$ conditions. A possible spiral, appearing as a faint, streak-like structure in the polar cap region, was identified only in the improved REPPU simulations that incorporate the solar wind–magnetosphere–ionosphere coupling system (Figure 8). The ionospheric spiral-associated FACs did not appear in the SWMF with BATS-R-US model (Figure S8 in Supporting Information S1). Although the REPPU simulations were performed with a $6^{th}$-order grid subdivision, corresponding to a spatial resolution of 1° in latitude and longitude directions, higher-order subdivision may be necessary for capturing the spiral in the polar cap more clearly. According to the SWMF with BATS-R-US simulation results (Figure 6; Figures S2 and S4 in Supporting Information S1), the magnetotail FAC intensity associated with the spiral was approximately three orders of magnitude weaker than that associated with the TPA, even though both the Polar observations (Figure 4) and the SWMF with BATS-R-US simulations show that the spiral source region was elongated tailward. This result suggests that the spiral extent alone does not necessarily determine the FAC intensity. In the REPPU simulations, the spiral-associated FAC may be faintly reproduced in the ionosphere, whereas the TPA-associated FAC system exhibits more coherent polarity and spatial correspondence between the magnetosphere and the ionosphere (Figure 8; Figure S7 in Supporting Information S1), suggesting that the TPA-associated FAC system is present but weakly developed. Such a difference likely reflects limitation in the current model resolution and/or in the representation of local-scale magnetotail process, rather than a fundamental inconsistency between the two models.

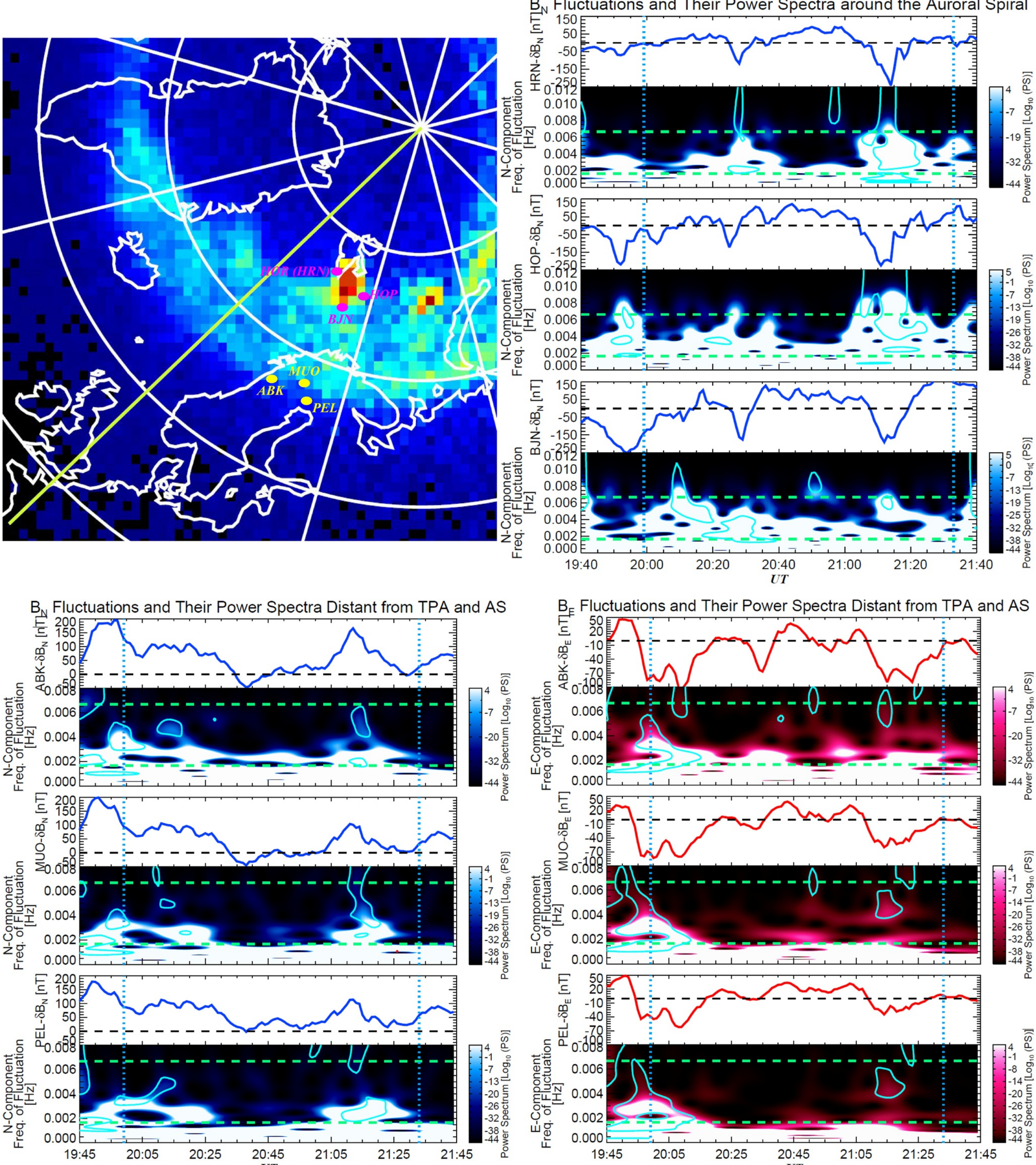

**Figure 11.** The locations of six ground magnetometer stations near (magenta) and equatorward (yellow; the northern part of the Scandinavia Peninsula) of the auroral spiral projected onto the Polar UVI image are shown in the left upper panel. Deviations of the $B_N$ component and their power spectrograms at HOR(HRN), HOP, and BJN stations during the interval when the auroral spiral was over and near the Svalbard from 19:59 UT to 21:33 UT, bracketed with two blue broken lines, are shown in the upper right panels. The $B_N$ and $B_E$ deviations and their power spectrograms at three geomagnetic observatories located equatorward (southward) of the spiral, corresponding to the northern part of the Scandinavia Peninsula, are shown in the bottom left and right panels. In the power spectrograms, the confidence level higher than 93% for the wave power intensity is surrounded by cyan solid ovals. The frequency range of the Pc5 waves (1.67–6.67 mHz) is bracketed by two green horizontal broken lines.

Based on the ground geomagnetic field perturbations, clear upward (from the ionosphere to the magnetosphere) FACs flew in association with the auroral spiral (Figure 9). To allow very weak FACs from the magnetotail to reach the polar cap effectively and form an auroral spiral, a solar wind-magnetosphere-ionosphere coupling system with minimal or no significant substorm effects appears to be necessary. One possible interpretation is that, even though the magnetotail spiral-associated FACs are generally weak, they may be sufficient to brighten the spiral in the ionosphere if the FAC sheets exhibit filamentary structures and are confined to a narrow region along the field lines during the formation or winding of the spiral. Nevertheless, previous studies using only local magnetosphere–ionosphere coupling models have also succeeded in reproducing auroral spirals (Lysak and Song, 1996; Partamies et al., 2001b).

Depending on the substorm phases, the corresponding magnetotail dynamics, including magnetic field and plasma variations, vary substantially. Therefore, it is natural that the formation process of the auroral spiral(s) during the substorm expansion phase, as addressed within an MHD regime (e.g., shear flow ballooning instability) and as examined by Keiling, Angelopoulos, Weygand, et al. (2009) and Voronkov et al. (2000), cannot be directly applied to our auroral spiral during the late stage of a substorm recovery phase. To gain insight into why the magnetotail FAC intensity associated with the spiral was so weak, the simulation results in Figure 7 are informative. According to the meridional magnetic field vector and FAC distributions in the distant magnetotail reproduced by the SWMF with BATS-R-US simulations (Figure S6 in Supporting Information S1), very weak spiral-associated FACs can be induced by flow shear between fast plasma flows driven by a plausible distant magnetotail reconnection at $X = \sim -133\ R_E$ and the background slower plasma. Figure 7a also shows that the source region of the spiral-associated FACs is characterized by relatively enhanced flow vorticity, which is consistent with the presence of reconnection-driven fast flows interacting with the surrounding slower plasma. This interpretation suggests that the closed field line region connected to the auroral spiral extends the distant magnetotail, consistent with a reconnection site located beyond $X = \sim -120\ R_E$. Regions of relatively stronger vorticity (flow shear) were elongated tailward in the dawnside TPA source region, compared to the duskside auroral spiral source region. Furthermore, the TPA source regions exhibited higher plasma densities and weaker magnetic field strengths than the auroral spiral source regions.

In this case, we consider only the FAC component that can be generated by the total time derivative of the magnetic field and the plasma vorticity driven by the plasma flow shear (the second term of the right-hand side of Equation 1, including Equations 3 And 4). The equations that describe total FAC components (e.g., Hasegawa & Sato, 1979; Sato and Iijima, 1979; Yao et al., 2012) are as follows:

$$J_{\parallel} = B \int \left\{ \frac{2J_{\perp} \cdot \nabla B}{B^2} + \frac{eN}{B} \frac{d}{dt} \left( \frac{\Omega}{\omega_c} \right) - \frac{J_{\mathrm{inertia}} \cdot e\nabla N}{e\mathrm{NB}} \right\} dl_{\parallel}, \tag{1}$$

$$J_{\perp} = J_{\mathrm{diamag}} + J_{\mathrm{inertia}}, \tag{2}$$

$$\Omega = (\nabla \times V) \cdot \hat{b}, \tag{3}$$

$$\frac{d}{dt} = \frac{\partial}{\partial t} + (V \cdot \nabla), \tag{4}$$

$$\omega_c = \frac{eB}{m}. \tag{5}$$

Here, $B$, $N$, $V$, $\Omega$, and $\hat{\boldsymbol{b}}$ in Equations 1–5 denote the magnetic field strength, the plasma density, the plasma bulk velocity, the plasma vorticity, and the unit vector of the magnetic field, respectively. From a series of Equations 1–5, ignoring the current density perpendicular to the magnetic field, comprising the diamagnetic ($J_{\mathrm{diamag}}$) and inertial currents ($J_{\text{intertia}}$) that are primarily produced by the pressure gradients and the plasma accelerations/decelerations, FACs can be described as follows (Hasegawa & Sato, 1979):

$$J_{\|} \approx B \int \frac{mN}{B} \frac{d}{dt}\left(\frac{\Omega}{B}\right) dl_{\|}. \tag{6}$$

According to Equation 6, the FACs are expected to be enhanced when the field strength is small, the associated plasma density is higher, and the vorticity is strong. Therefore, spiral-associated FACs are likely to be weaker than those in the TPA source region, because the spiral source regions exhibit lower densities and stronger magnetic field intensities, despite the locally intense vorticities in the spiral source region (Figures 6c and 7a). This simple consideration also supports that a mechanism and a process are required to form the spiral based on the weak magnetotail FACs. For a future study, it is necessary to perform spiral-specified global MHD simulations with higher temporal and spatial resolutions, that is, simulations using the improved REPPU code with higher order number (7th and 8th) grid subdivision. We are expecting to obtain some new insights for the auroral spiral that is taken into account the solar wind-magnetosphere-ionosphere coupling system with minimal or no significant substorm effects.

Ground-based geomagnetic field measurements in the equatorward (southward) of the spiral clearly detected ULF waves in the Pc5 range. The solar wind dynamic pressure during the auroral spiral appearance (Figure 2) shows a gradual enhancement. It remains unclear whether such a change is sufficient to excite the ULF waves associated with the spiral formation. Nevertheless, the observed ULF wave activity may indicate a contribution to particle energization processes involved in the spiral formation, such as auroral particle acceleration via field line resonance (FLR; Rankin et al., 2005). It still remains open how these observed ULF waves are related to local-scale auroral spiral formation and how they are triggered, if these waves are FLR signatures. These interesting questions are also future work to be resolved.

## Conflict of Interest

The authors declare no conflicts of interest relevant to this study.

## Availability Statement

Polar ultraviolet imager (UVI) level-1 data can be accessed from https://cdaweb.gsfc.nasa.gov/pub/data/polar/uvi/uvi_level1/. Data for calibrating the level-1 data and calculating the position of the UVI images and Polar UVI data calibration toolkit (software) can be accessed from Uritsky and POLAR UVI team (2017). All IMAGE magnetometer numerical data used in Figure 10 can be downloaded from the IMAGE website (https://space.fmi.fi/image/www/?page=user_defined). We also gratefully acknowledge the SuperMAG collaborators (https://supermag.jhuapl.edu/info/?page=acknowledgement) for using the *SML* and *SMU* indices. The numerical data from the magnetometers used in Figure 11 as well as the *SML* and *SMU* indices can be downloaded from the SuperMAG website (https://supermag.jhuapl.edu/mag/). The OMNI solar wind magnetic field and plasma data can be acquired from Coordinated Data Analysis Web (https://cdaweb.gsfc.nasa.gov/pub/data/omni/omni_cdaweb/hro_1min/1997/), which is administrated by GSFC/NASA. The $K_p$ index is provided by the Helmholtz Centre Potsdam - GFZ German Research Centre for Geosciences (https://kp.gfz.de/en/). The SWMF with BATS-R-US MHD simulation results can be referred to the run of Motoharu_Nowada_090523_2 (https://ccmc.gsfc.nasa.gov/ror/results/viewrun.php?domain=GM&runnumber=Motoharu_Nowada_090523_2; https://ccmc.gsfc.nasa.gov/cgi-bin/vis3d/CCMCVis_3D.cgi?dir=33478; https://ccmc.gsfc.nasa.gov/cgi-bin/CCMCVis_2D.cgi?dir=33478) in the CCMC. All numerical data of the REPPU global MHD simulation results in this study are available in Nowada et al. (2024).

**Acknowledgments**

M.N. enjoyed fruitful and constructive discussions with Qiu-Gang Zong, Alexander William Degeling, Timo Pitkänen, and Jong-Sun Park, and was supported by a Grant of the National High-end Foreign Experts Introduction Program (H20240313), sponsored by the State Administration of Foreign Experts Affairs (SAFEA)/Ministry of Science and Technology (MOST). Y.M. was supported by basic research funding from Korea Astronomy and Space Science Institute (KASI2026183005). Q.Q.S. was supported by NSFC 41731068, 41961130382, and 41974189. We thank George K. Parks for providing the Polar UVI data and Kan Liou for processing the data. Simulation results have been provided by the Community Coordinated Modeling Center (CCMC) at NASA Goddard Space Flight Center through their publicly available simulation services. This work was carried out using the SWMF and BATS-R-US tools developed at the University of Michigan's Center for Space Environment Modeling (CSEM). The modeling tools described in this publication are available online through the University of Michigan and are also accessible via the CCMC. We thank the institutes that maintain the IMAGE Magnetometer Array: Tromsø Geophysical Observatory of UiT, the Arctic University of Norway (Norway), Finnish Meteorological Institute (Finland), Institute of Geophysics, Polish Academy of Sciences (Poland), GFZ German Research Centre for Geosciences (Germany), Geological Survey of Sweden (Sweden), Swedish Institute of Space Physics (Sweden), Sodankylä Geophysical Observatory of the University of Oulu (Finland), Polar Geophysical Institute (Russia), DTU Technical University of Denmark (Denmark), and Science Institute of the University of Iceland (Iceland).

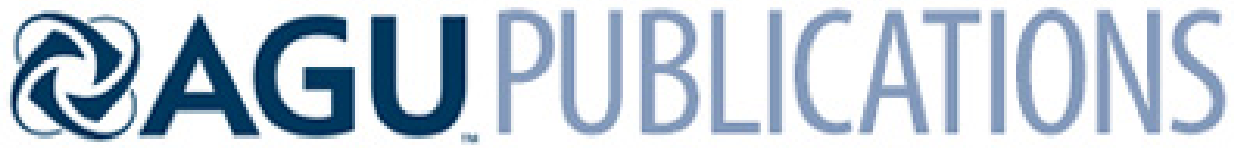

# Contemporaneous Appearances of Auroral Spiral and Transpolar Arc: Polar UVI Observations and Global MHD Simulations

Motoharu Nowada[1]*, Yukinaga Miyashita[2,3], Aoi Nakamizo[4],
Noora Partamies[5], and Quan-Qi Shi[1]

1: Shandong Key Laboratory of Optical Astronomy and Solar-Terrestrial Environment, School of Space Science and Physics, Institute of Space Sciences, Shandong University, Weihai, Shandong, People's Republic of China.
2: Center for Heliophysics Research, Korea Astronomy and Space Science Institute, Daejeon, South Korea.
3: Department of Astronomy and Space Science, Korea University of Science and Technology, Daejeon, South Korea.
4: National Institute of Communications and Technology, Koganei, Tokyo, Japan.
5: Department of Arctic Geophysics, The University Centre in Svalbard, Norway.

*Corresponding author: Motoharu Nowada (moto.nowada@sdu.edu.cn; nowada.sdu@gmail.com)

## Contents of this file

## Introduction

Figure S1 presents overlaid plots of field-aligned currents (FACs; color scale), magnetic field vectors, and total pressure contours. It also includes panels showing total pressure or FACs (color scale), together with plasma velocity vectors, and magnetic field lines categorized as interplanetary magnetic field, closed, and polar-cap-originated field lines. These quantities are shown for the meridional and magnetotail cross sections at X = -5.0, -15.0, -30.0, and -45.0 $R_E$. Figures S2 and S4 show the spiral-associated magnetotail FAC distributions simulated using SWMF with BATS-R-US code at 20:16 and 21:08 UT, respectively. The associated plasma flow vorticity, plasma density, and magnetic field intensity distributions in the magnetotail at these times, obtained from the SWMF with BATS-R-US MHD simulations, are presented in Figures S3 and S5. Figure S6 shows meridional (X-Z plane) magnetic field vector plots overlaid on the FAC distributions, extending from the near-Earth to the distant magnetotail regions. This plot is taken along a cut at Y = 7.0 $R_E$, corresponding to the location of the auroral spiral. Figure S7 shows the magnetotail FAC structures associated with the TPA and the auroral spiral reproduced by the improved REPPU simulations. In the improved REPPU model, the TPA- and spiral-associated FAC structures are projected onto the dipole equatorial plane (X-Y plane in Solar Magnetic (SM) coordinates), because the code is formulated exclusively in the SM coordinate system. Figure S8 presents the ionospheric FAC distributions during contemporaneous appearances of the TPA and the auroral spiral, as reproduced by SWMF with BATS-R-US MHD simulations. Figure S9 presents deviations of the $B_E$ and $B_Z$ components observed at geomagnetic observatories located near, poleward (northward), equatorward (southward), and westward of the auroral spiral from 19:45 to 21:45 UT, which are not discussed in the main text. Figure S10 shows deviations of the $B_N$ and $B_Z$ components and their power spectrograms at the HOR(HRN), HOP, and BJN stations which are located in and near Svalbard. Figure S11 presents the plots of the $dB_Z$ component and its power spectrogram at ABK, MUO, and PEL, corresponding to the locations equatorward (southward) of the spiral during contemporaneous appearances of the TPA and the auroral spiral.

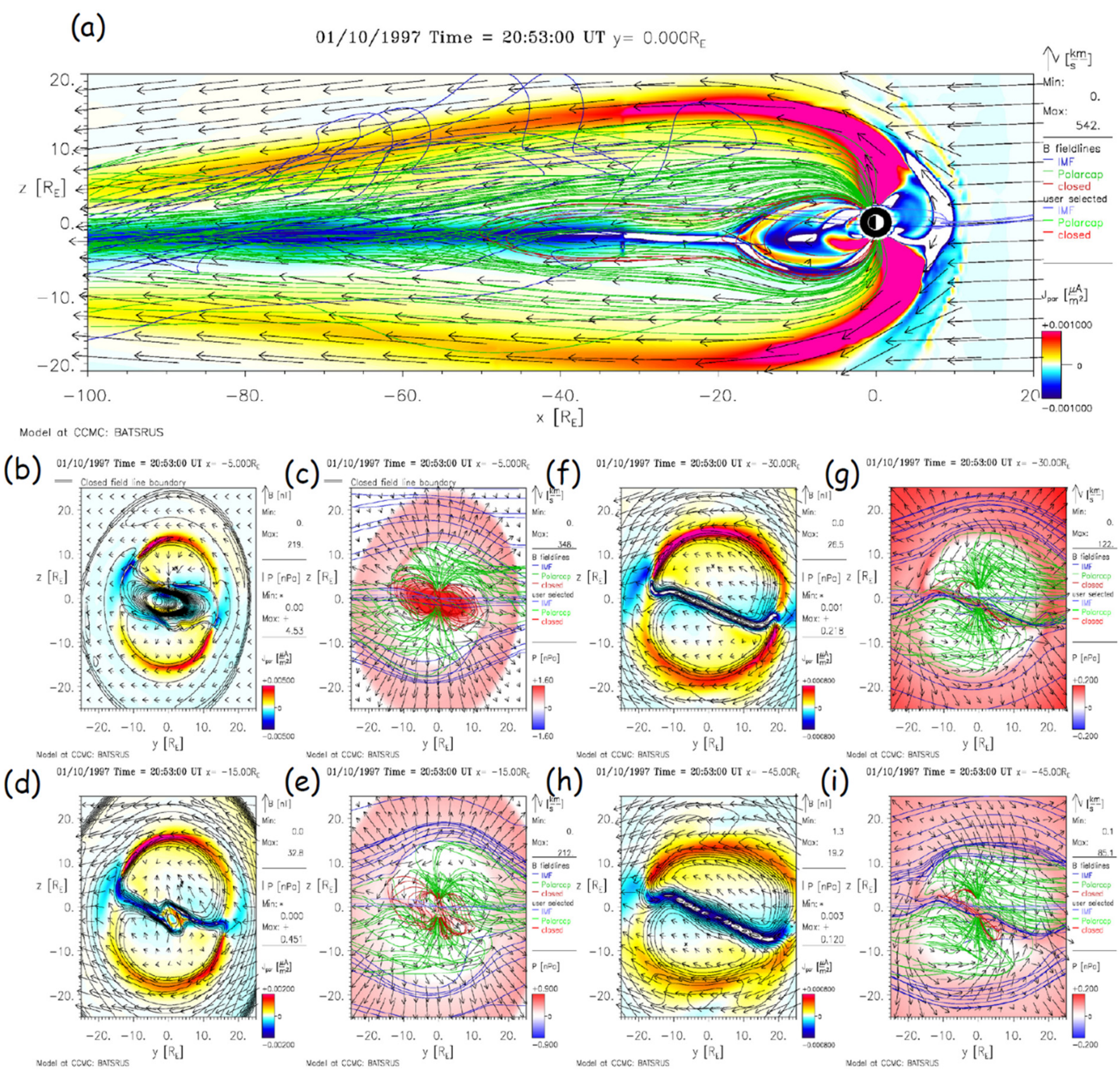

**Figure S1.** Meridional (panel a) and cross-sections of the magnetotail region at X = -5.0 $R_E$ (panels b and b), X = -15.0 $R_E$ (panels d and e), X = -30.0 $R_E$ (panels f and g), and X = -45.0 $R_E$ (panels h and i), covering the range from -25 to 25 $R_E$ in X-Z plane, are shown. The time corresponds to that in Figure 6, when the TPA and the auroral spiral were contemporaneously observed by Polar UVI (20:53 UT). In each pair of panels, the left (top) and right (bottom) sides correspond to the dawnward (northward) and duskward (southward) directions along the Y(Z) axis, respectively. Panel (a) shows overlaid plots of FAC ($J_{//}$, colors), plasma velocity (vectors), and magnetic field lines categorized as IMF (blue), closed (red), and polar-cap-originated (green) field lines. In panels (b), (d), (f), and (h), superimposed plots of FAC ($J_{//}$, colors), magnetic field line (vectors), and total pressure (contours) are presented. Panels (c), (e), (g), and (i) show overlaid plots of total pressure (red-blue color scale), plasma velocity (vectors), and three-categorized magnetic field lines.

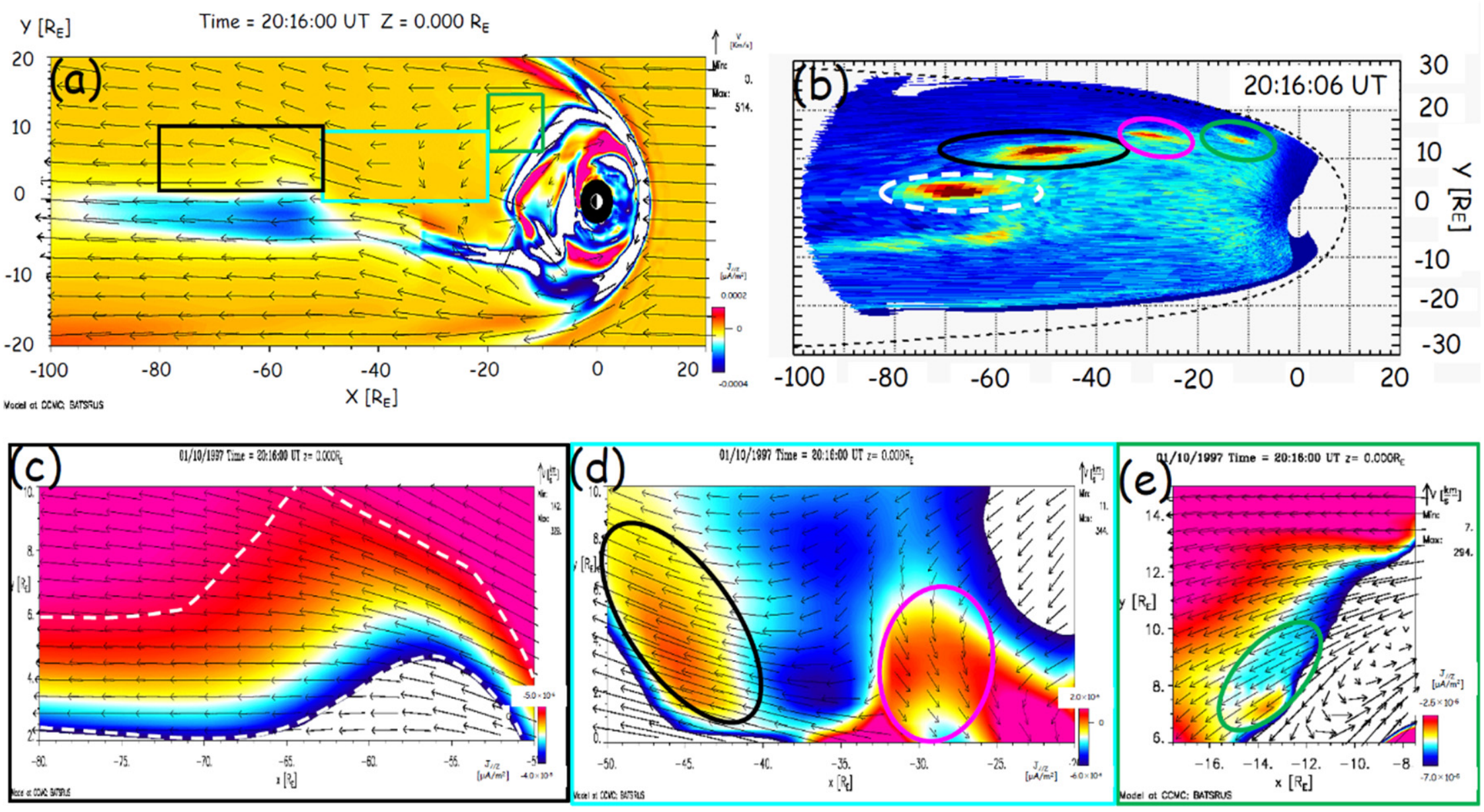

**Figure S2.** An example of comparison between magnetotail FAC distributions reproduced by the SWMF with BATS-R-US global MHD simulations (panel S2a) and observational results (20:16:06 UT, panel S1b), and zoomed-in plots of simulated FACs ($J_{//Z}$, panels S2c – S2e) are presented. The format of the color codes and figures are the same as those in Figure 6. The color code scales of panels S2(c) – S2(e) are different from that in panel S2(a), and those in panels S2(c) and S2(e) indicate the downward FAC intensity. The black, cyan, and green squares in panel S2(a) correspond to the regions shown in panels S2(c) – S2(e). The regions surrounded by green, magenta, and black ovals, and broken white curves in panels S2(c) – S2(e) correspond to the regions highlighted with the ovals in the same colors in panel S2(b).

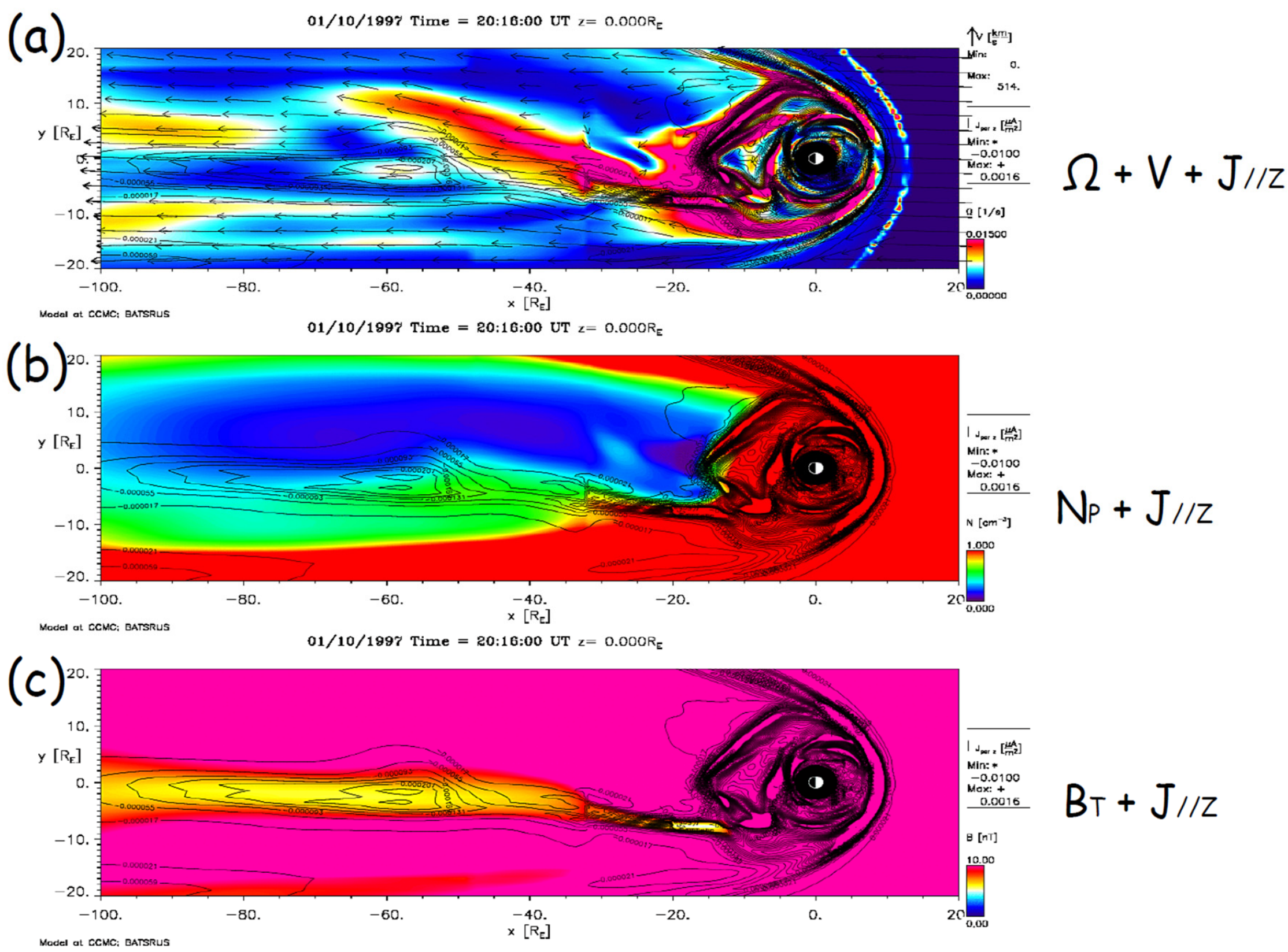

**Figure S3.** The SWMF with BATS-R-US MHD simulation results of the vorticity (panel S3a), plasma density (panel S3b), and magnetic intensity (panel S3c) distributions on the magnetotail equatorial plane on 20:16:00 UT are shown. The color code formats are the same as those in Figure 7.

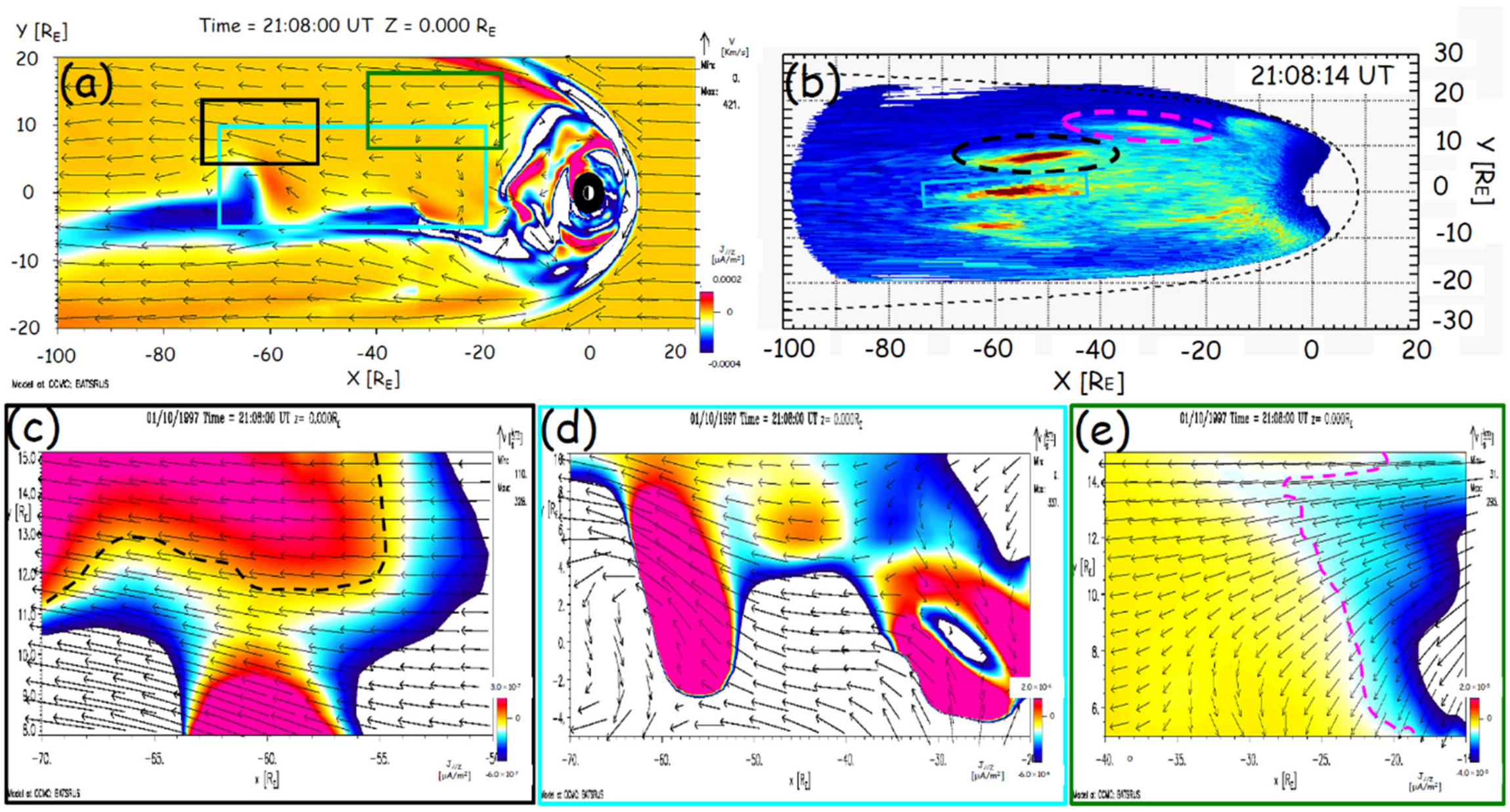

**Figure S4.** Another example of comparison between magnetotail FAC distributions reproduced by the SWMF with BATS-R-US global MHD simulations (panel S4a) and observational results (21:08:00 UT, panel S4b), and zoomed-in plots of simulated FACs ($J_{//Z}$, panels S4c – S4e) are presented. The format of the color codes and figures are the same as those in Figures 6 and S2. The color code scales of panels S4(c) – S4(e) are different from that in panel S4(a). The black, cyan, and green squares in panel S4(a) correspond to the regions shown in panels S4(b) – S4(e). The regions surrounded by broken black and magenta curves in panels S4(c) and S4(e) correspond to the regions highlighted with the ovals in the same colors in panel S4(b).

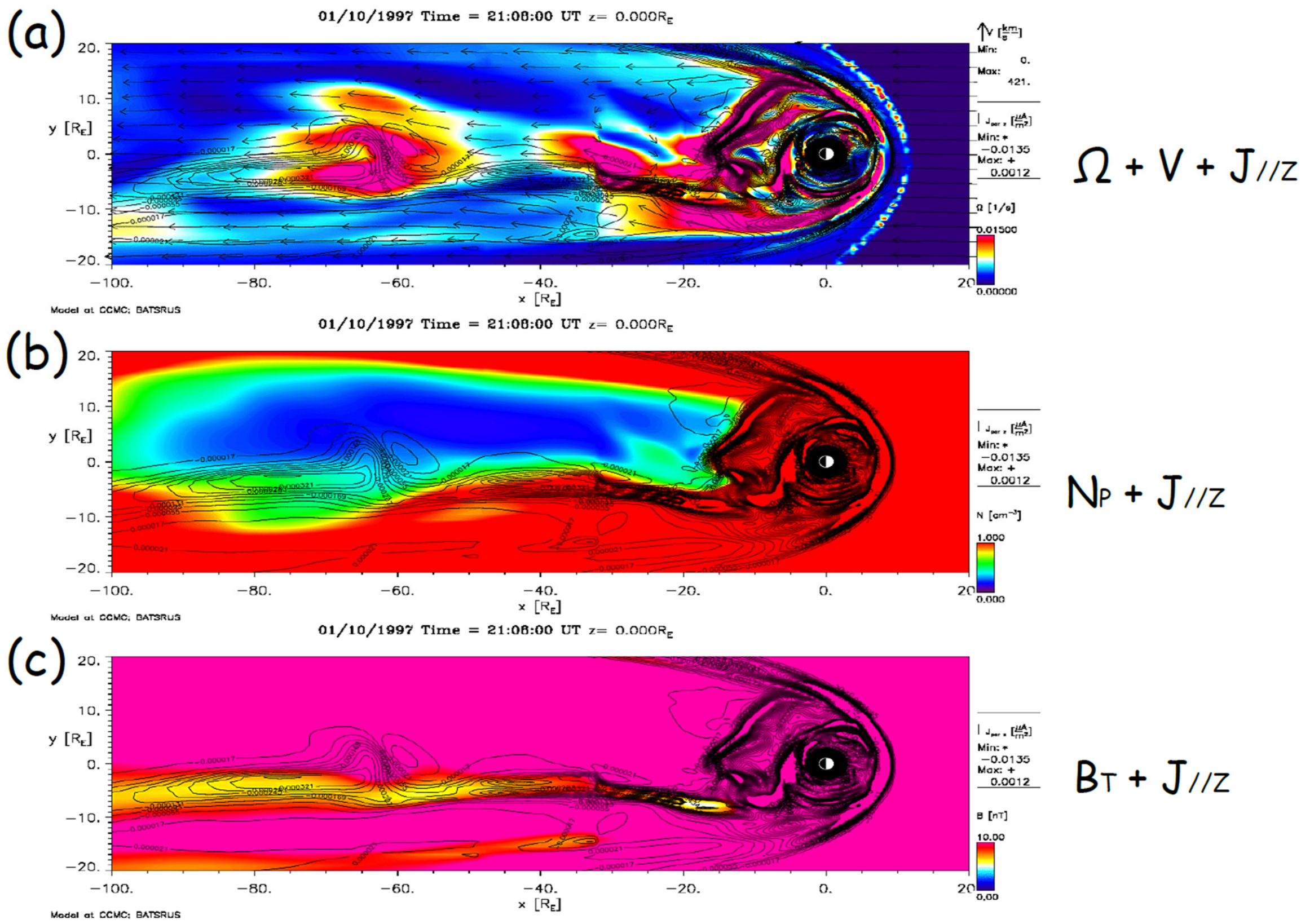

**Figure S5.** The SWMF with BATS-R-US MHD simulation results of the vorticity (panel S5a), plasma density (panel S5b), and magnetic intensity (panel S5c) distributions on the magnetotail equatorial plane on 21:08:00 UT are shown. The color code formats are the same as those in Figures 7 and S3.

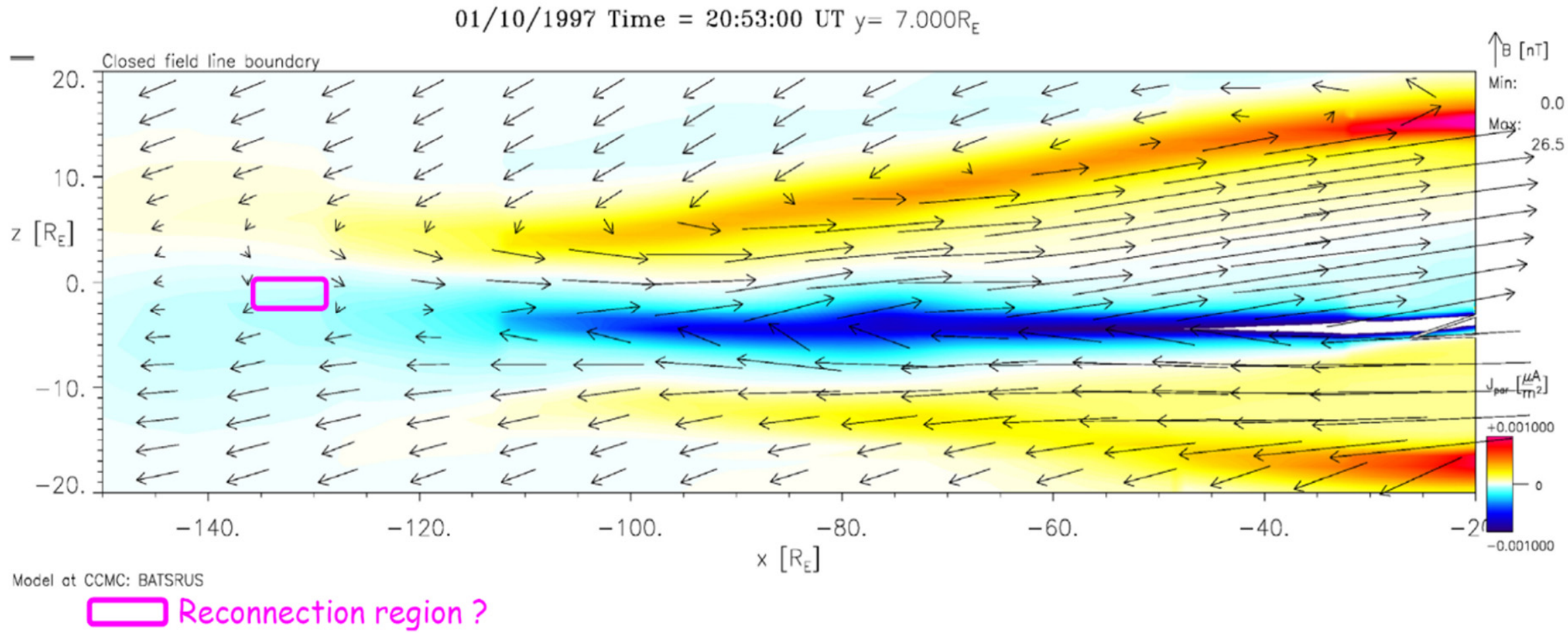

**Figure S6.** Meridional (X-Z plane) magnetic field vector plots overlaid on the FAC distributions, represented by color codes, are shown from the near-Earth to the distant magnetotail regions. The plot is taken along a cut at Y = 7.0 R$_E$ corresponding to the location of the auroral spiral identified in Figure 6(d). A plausible reconnection region, identified by the field line reversal, is highlighted by the magenta square.

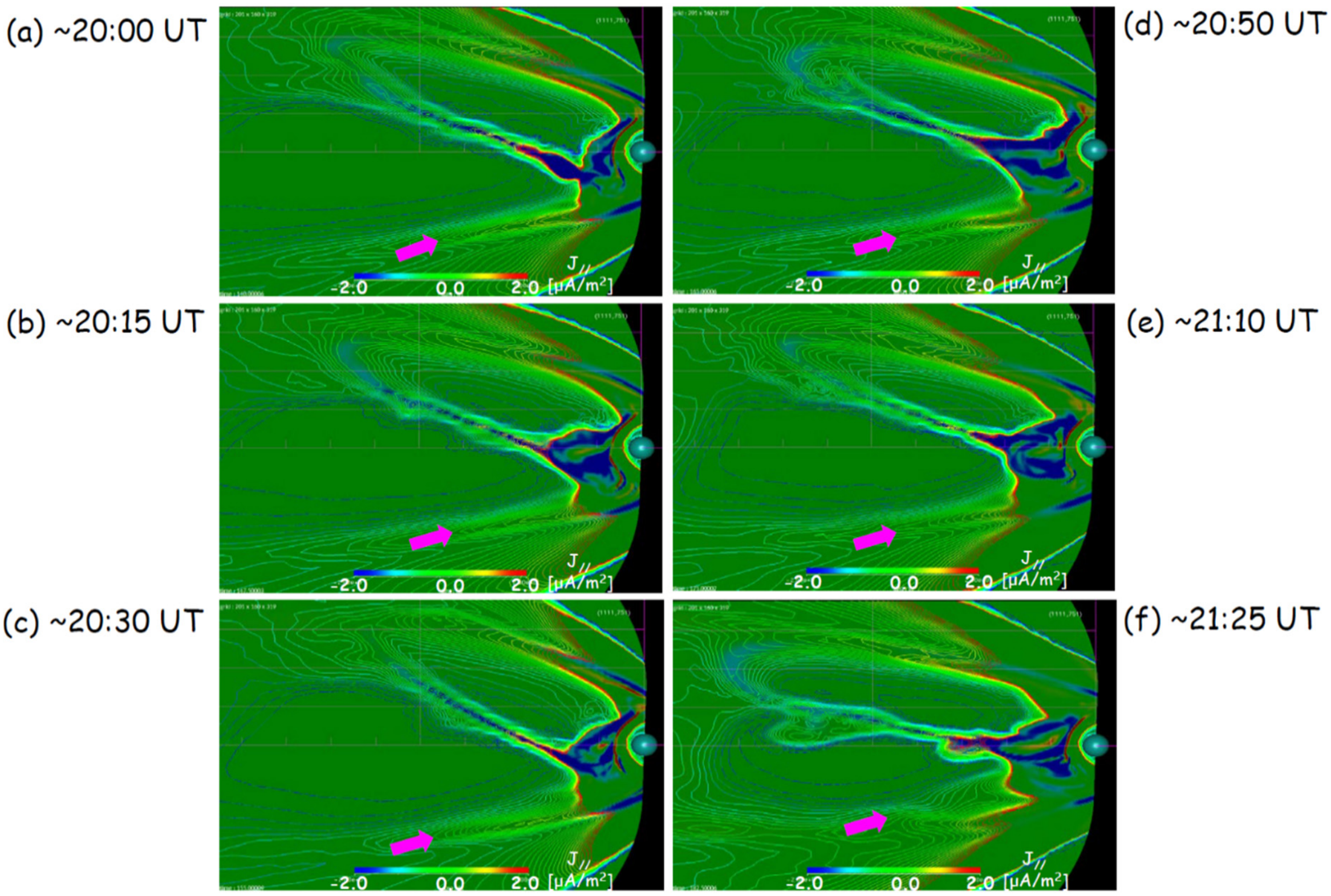

**Figure S7.** The FAC ($J_{//}$) structures projected onto the X-Y plane in Solar Magnetic (SM) coordinates (dipole equatorial plane) in the Northern Hemisphere are shown. The color codes show the orientation and intensity of the FACs. The upward (directed out of the plane) and downward (into the plane) FACs are shown bluish and reddish colors. The TPA source regions identified are indicated with the thick magenta arrows in each panel. The time stamps from panels S7(a) to S7(f) are almost fixed to those in Figure 5.

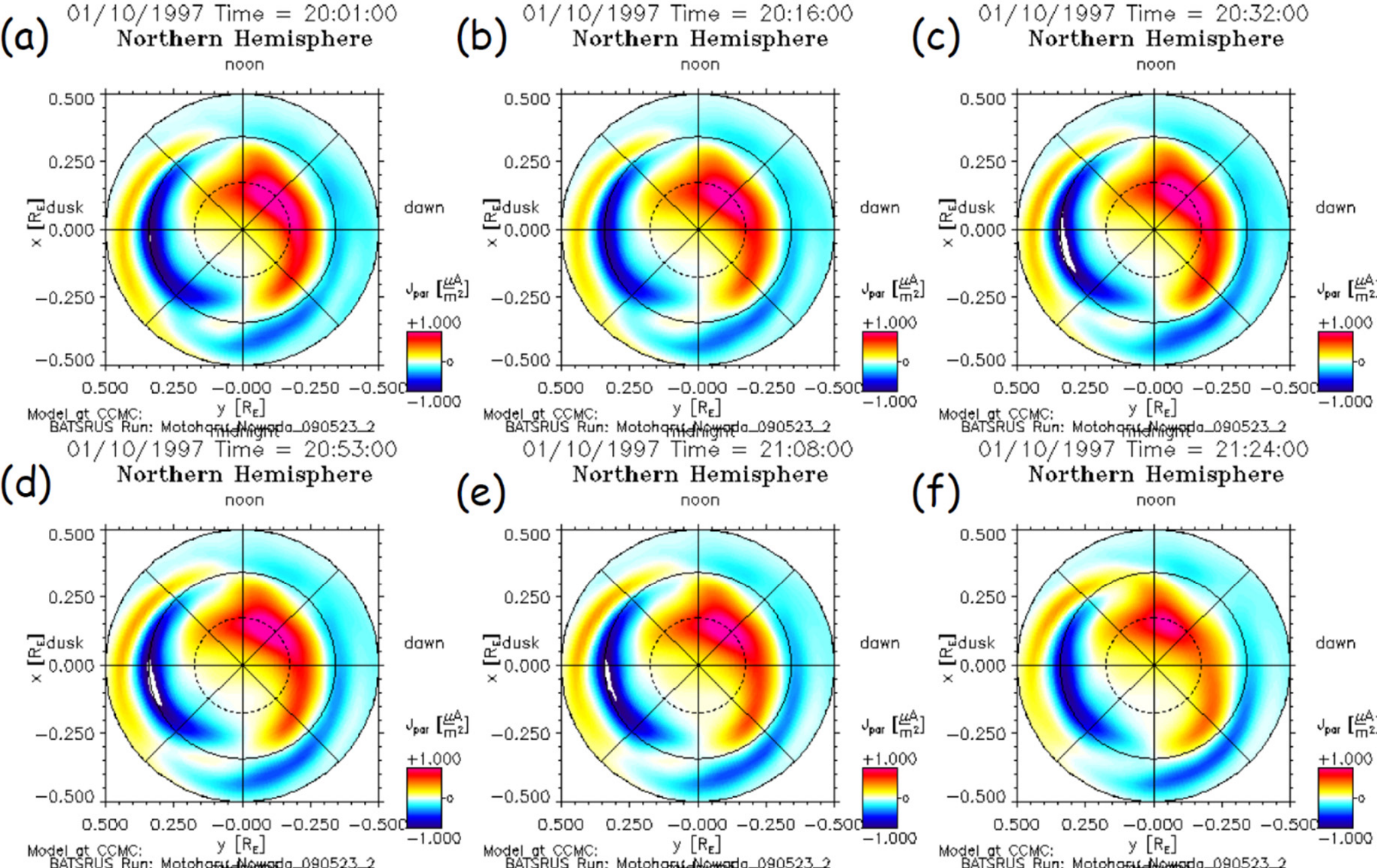

**Figure S8.** Ionospheric FAC ($J_{//}$) distributions during contemporaneous appearances of the TPA and the auroral spiral in the Northern Hemisphere, as observed by Polar UVI and reproduced by global MHD simulations using the SWMF with BATS-R-US model, are presented. The timestamps correspond to those in Figures 4 and 5. Color scales are assigned according to the FAC intensity and orientation, where reddish (bluish) colors indicate upward/ionosphere-to-magnetosphere (downward/magnetosphere-to-ionosphere) currents. Black solid and dotted circles denote magnetic latitudes from 60° to 80°, and radial straight lines indicate magnetic local time (MLT) at 3-h intervals. Each panel is oriented such that the bottom, right, top, and left correspond to midnight (0 h MLT), dawn (6 h MLT), noon (12 h MLT), and dusk (18 h MLT), respectively.

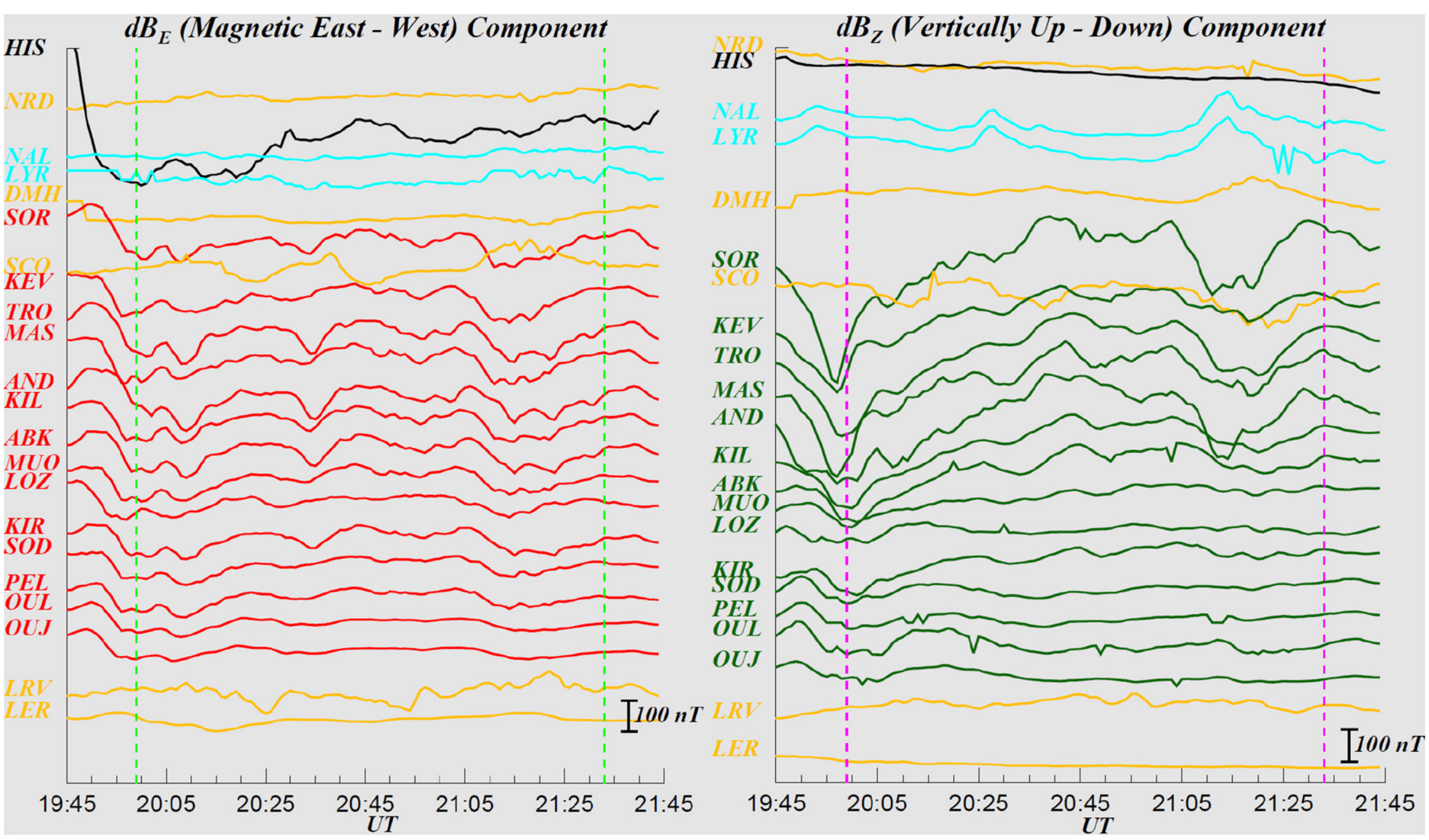

**Figure S9.** Deviations of the E (left panel) and Z (right panel) magnetic field components at the stations around the Svalbard during the auroral spiral interval from 19:45 UT to 21:45 UT are shown. The time interval when the TPA and the spiral were contemporaneously seen is bracketed with two green and magenta vertical broken lines.

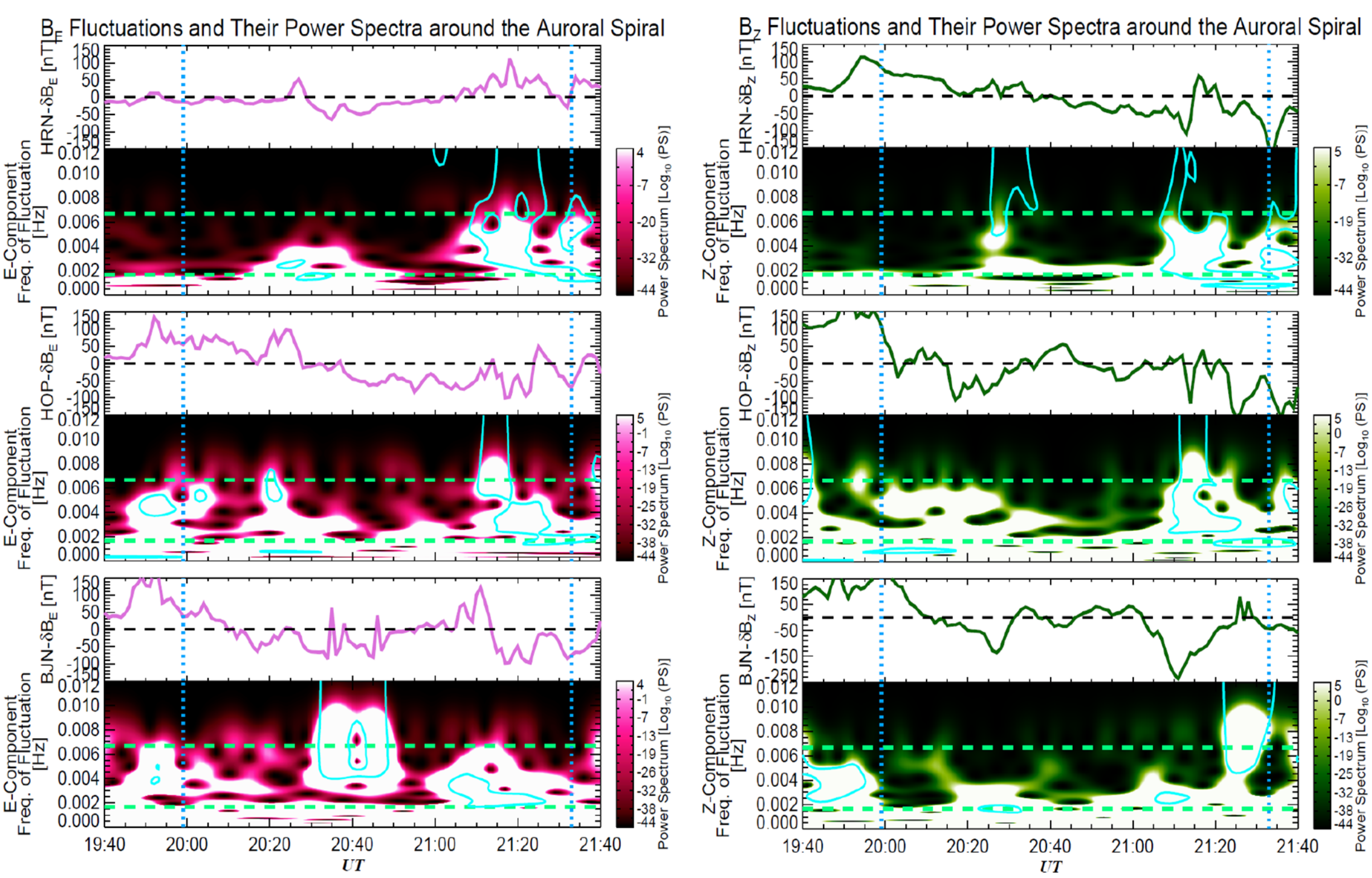

**Figure S10.** Deviations of the $B_N$ and $B_Z$ components and their power spectrograms at HOR(HRN), HOP, and BJN stations from 19:40 to 21:40 UT, including the time interval when the auroral spiral was over and near the Svalbard from 19:59 UT to 21:33 UT, are shown. Figure formats are the same those in the spectrograms presented in Figure 11.

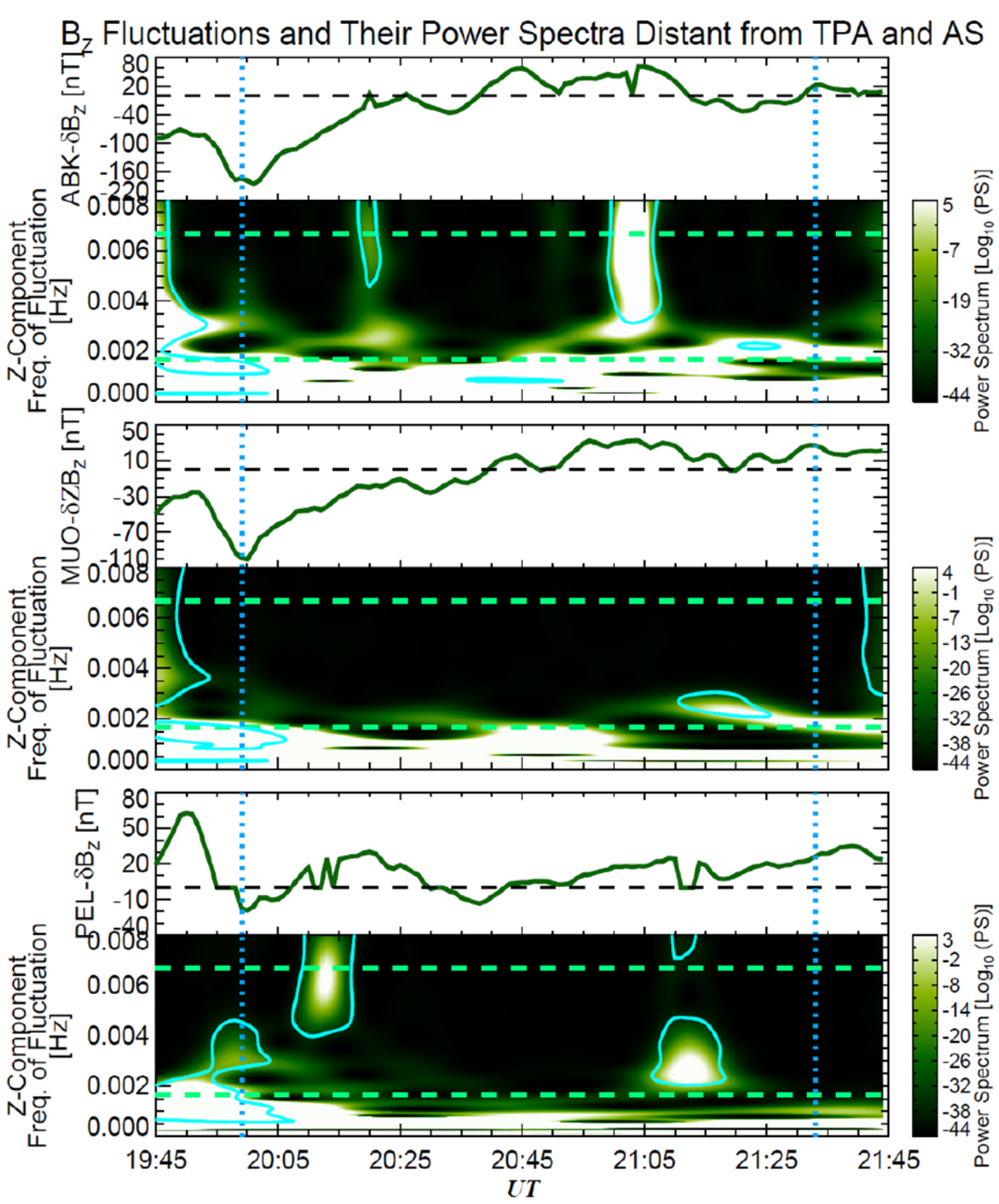

**Figure S11.** The $B_Z$ deviation and its power spectrogram at three geomagnetic observatories (ABK, MUO, and PEL), corresponding to the locations equatorward (southward) of the spiral, are shown. Figure format is the same as that in Figures 11 and S10.